\documentclass[11pt]{article}

\usepackage{a4wide}
\usepackage{amssymb}
\usepackage{amsmath}
\usepackage{amsthm}
\usepackage[utf8]{inputenc}
\usepackage{hyperref}
\usepackage{xcolor}

\usepackage{booktabs,longtable,array,ragged2e,pdflscape}
\usepackage{multirow}
\usepackage{pgfplots}
\pgfplotsset{width=0.9\linewidth,compat=1.17}
\usepgfplotslibrary{fillbetween} 
\usepackage{xcolor}

\begin{document}

\title{Choosing an energy-efficient software architecture for building system diagnostic support}

\author{Roxane Koitz-Hristov and Franz Wotawa \\
Graz University of Technology \\
Institute of Software Engineering and Artificial Intelligence \\
Inffeldgasse 16b/2, A-8010 Graz, Austria \\
\{rkoitz-hristov, wotawa\}@tugraz.at}

\date{}

\maketitle

\begin{abstract}
Around 30\% of global energy expenditure can be attributed to the building sector, where a large portion of energy-consumption could be avoided by repairing existing faults. Fault detection and diagnosis (FDD) software addresses this issue; however, its creation and operation also have an environmental impact. The magnitude of this impact is influenced by the diagnosis architecture, as different architectures and methods have different energy demands. Yet, simply considering the energy consumed by the software itself is not sufficient to assess its overall environmental impact, since the diagnostic performance, e.g., number of detected faults or number of faults missed, also contributes to its ecological footprint. 

In this paper, we propose an energy-consumption model that considers FDD performance and energy spend directly by the diagnosis software. In an initial experiment, we compare several FDD architecture families, i.e., rule-based, model-based, classical machine learning, and large-language-model-based, in simulation using performance and energy-consumption values collected from prior literature.   
The results show that considering the computational energy and accuracy of FDD can change the relative benefit of the different approaches. Computationally efficient machine learning methods, such as random forest, provide the largest net savings on smaller buildings, whereas more resource-intensive approaches, such as fine-tuned large language models, become advantageous as building size increases. Our findings suggest that overall energy efficiency depends not only on the computational demand of the FDD software, but also on its diagnostic performance and the scale of the building. 
\end{abstract}

\section{Introduction}
\label{sec:intro}


The energy expenditure of residential and commercial buildings constitutes a critical challenge in achieving the United Nations' climate targets. Recent estimates indicate that the domain is responsible for 40\% of the final energy-consumption in the European Union~\cite{EC2023} and  30\% worldwide~\cite{UNEP_Global_Status_Report_2024}. Roughly 75\% of the EU building stock is energy inefficient~\cite{EU_EnergyEfficiencyBuildings_2020}, with operational faults as a major contributing factor~\cite{DGNB}. Notably, heating, ventilation, and air conditioning (HVAC) systems account for a considerable portion of the energy demand in the building sector. Studies estimate that simply repairing existing HVAC faults could substantially reduce the energy-consumption~\cite{bellanco2021Review}. Thus, fault detection and diagnosis (FDD) for building systems is a highly relevant strategy for ensuring energy efficiency and the achievement of sustainability goals. 


Historically, FDD in buildings has primarily relied on knowledge-based approaches~\cite{kim2018review,cho2005transient,sterling2014model,koitzhristov_et_al:OASIcs.DX.2025.1}. However, in recent years, the widespread adoption of building automation systems (BAS), increasing availability of sensor data, and advancements in data science have led to the development of various machine learning (ML) techniques for FDD~\cite{zhou2023comprehensive}. ML has proven to be a suitable approach, which can achieve high detection and diagnostic accuracy while requiring little to no prior knowledge of the system's underlying composition or dynamics~\cite{Koitz-Hristov2026,chen2023review, zhang2023deep,IEA_Buildings_2022}.

One goal of FDD in buildings is to improve energy efficiency, yet sophisticated FDD methods can be computationally costly to train and operate, leading potentially to a substantial amount of energy use and therefore additional carbon emissions. The questions of how much energy software requires and which design choices influence its resource consumption have been studied in software engineering previously. In software engineering, sustainability has been established as a multi-dimensional property of software quality in its own right, which includes technical, economic, social, and environmental aspects~\cite{Lago15,Verdecchia21,calero2015green}. In our work, we are interested in the environmental dimension, in particular, in the environmental impact of FDD software. 
While most definitely the implementation itself, i.e., the underlying code, affects the ecological footprint of software, there have been studies that show that also software architecture can measurably influence the energy demands of a system~\cite{Fatima,PROCACCIANTI20152,Jagroep2017}. In general, architectural decisions shape a system's structure and behavior. Therefore they influence the quality and sustainability of software products. Moreover, changing design decisions is expensive, especially later in the software life cycle~\cite{VENTERS2018174}. 

With the adoption of complex ML methods, which are associated with high energy demands for training~\cite{strubell2019energypolicyconsiderationsdeep}, the conversation on software sustainability has extended to the domain of artificial intelligence (AI). A recent debate contrasts ``Red AI'', which prioritizes performance without regard for environmental cost, with ``Green AI'', which favors computational efficiency and reduced energy use while maintaining acceptable accuracy~\cite{Schwartz20,verdecchia2023systematic}. Recent work addresses the architectural angle within the discussion around sustainable ML with the goal to determine ``green'' architectural strategies~\cite{martinez23,Jarvenpaa24}.

\begin{sloppypar}
``Green software'' and ``Green AI'' often aim to reduce the energy-consumption of the software under consideration. In the building domain, the FDD software not only consumes energy itself, but it is also intended to enable energy savings in the building it monitors. Thus, to minimize the overall environmental footprint, one has to consider the energy consumed by the software as well as the energy savings enabled by the software. In practice, this raises the question of which of the available FDD techniques is the most overall energy-efficient solution. Answering this question is more intricate than it initially seems, since the commonly used methods differ largely in the computational resources they require and their diagnostic quality. For instance, diagnosing a building from a small set of threshold-based rules can be done very efficiently, however, if the rule-based system raises a large number of false alarms the additional energy needed for the unnecessary maintenance activities offsets some of the computational efficiency of the FDD approach. The reverse may also happen, where a model is associated with substantially more resources to be able to diagnose a system, e.g., a deep neural network, yet it may be able to detect faults more reliably and therefore may decrease the energy expenditure of the building. Consequently, the computational energy required to operate the diagnosis software has a direct environmental impact, whereas the energy savings resulting from diagnosis activities have an indirect, enabling impact~\cite{Lago19}. Often there might be a tradeoff between diagnostic performance and environmental impact, yet both affect the ecological footprint of a building. Thus, in order to answer the question which FDD technique is the most overall energy-efficient solution, we must analyze which combination of the two factors maximizes energy savings. 
\end{sloppypar}

In this paper, we address this question in the context of building HVAC systems by making the following contributions: First, we introduce an energy-consumption model that relates fault detection performance of an FDD method to the energy-consumption of a building affected by faults. Second, we extend this model with energy requirements associated with the training of the diagnosis method as well as its inference. Third, we compile diagnosis performance and energy-consumption data reported for the considered methods from the FDD literature and use the resulting values to compare the methods in simulation experiments to assess the suitability of the different diagnosis architectures and methods. 




\section{Diagnosis System Architecture}\label{sec:diagarch}


In this section, we discuss the different FDD methods considered in this study together with their corresponding high-level software architectures. Afterward, we summarize different evaluation metrics for FDD, which we later use to estimate the energy consumption of the introduced methods. 

\subsection{Diagnosis methods}

{\em Rule-based systems} have been successfully used in medicine~\cite{buc84} but also in diagnosis of engineered systems~\cite{ret87,rie90}. The underlying idea of (rule-based) expert systems is to encode domain expertise as statements such as {\tt if} $observation$ {\tt then} $cause$ to determine the reason behind a certain observation or symptom. 

\begin{figure}
\centering
\includegraphics[width=0.8\linewidth]{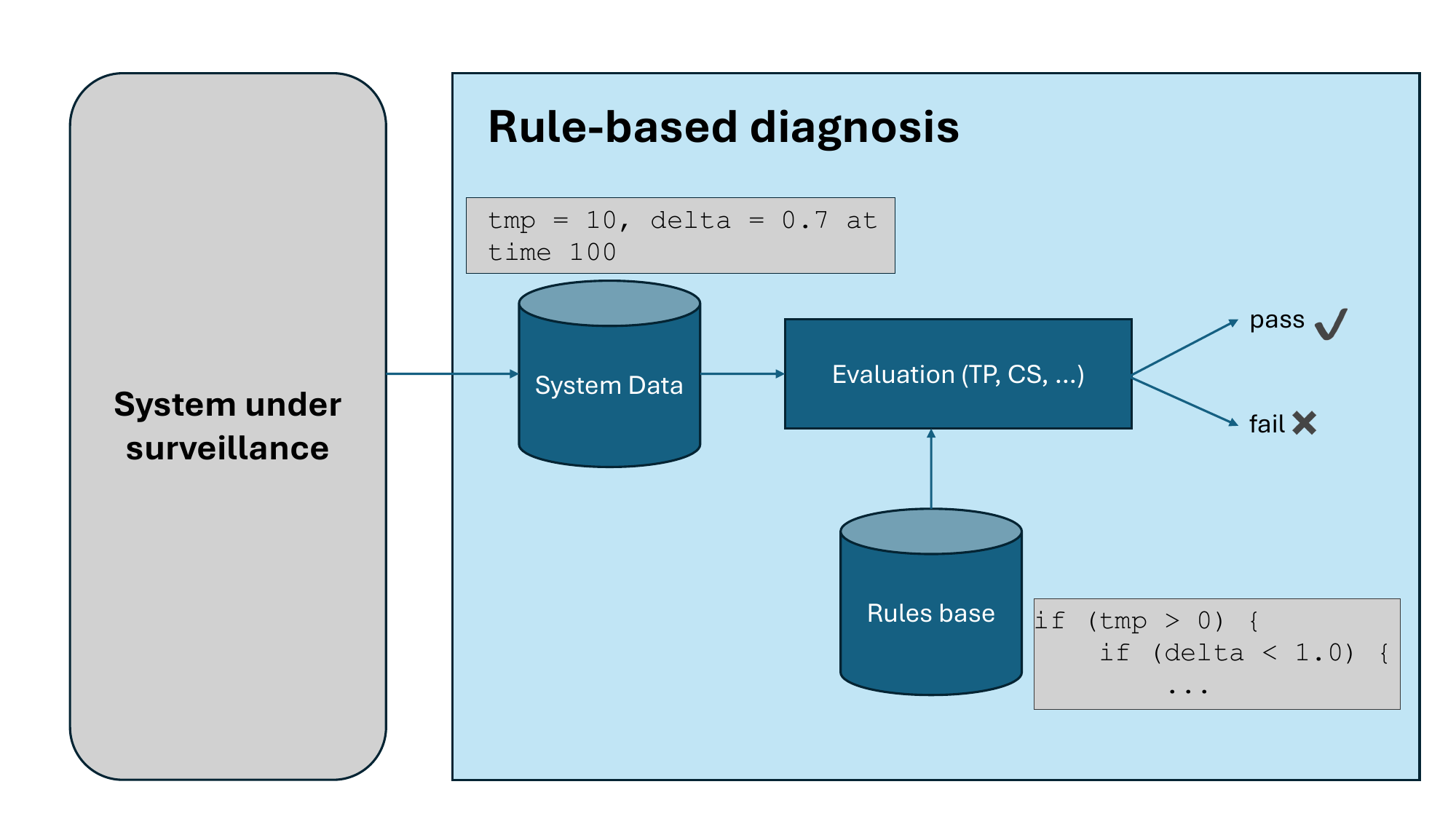}
\caption{Rule-based diagnosis architecture}\label{fig:rbarch}
\end{figure}

Figure~\ref{fig:rbarch} depicts the general architecture of an FDD system utilizing rules for detecting faults. There is a component for fetching observations from the system under surveillance, i.e., a monitoring system~\cite{DBLP:conf/seke/StettingerW23}, which stores the information in a system data base. 
The evaluation component checks for consistency regarding the given rules, which are stored in a rule base. This evaluation component can be as simple as any ordinary program implementing {\tt if-then-else} rules or a more sophisticated theorem prover that checks whether the current data is consistent with the rules (see, e.g.,~\cite{DBLP:conf/dsa/KlampflPW25,DBLP:conf/dx/KlampflW24}). Depending on the underlying detailed architecture, i.e., whether the rule-based diagnosis system relies on simple programs or a theorem prover, the energy-consumption varies. However, the quality of the diagnostic result should not be influenced by details regarding the concrete implementation. 

Although rule-based systems can encompass a wide variety of knowledge, the interdependence between rules makes this type of system difficult to maintain as the size of the rule base increases.
As an answer to this limitation, researchers have proposed {\em model-based diagnosis}~\cite{dav82,dav84,rei87,kle87}. Model-based diagnosis differs from rule-based diagnosis in the way faults are detected. The former utilizes a model for detecting a faulty behavior, which requires means for executing the model in order to derive the expected system behavior from the model. The expected behavior is then compared with the observed behavior. Hence, inference is more complex compared to ordinary rule-based systems. Therefore, each inference step has a higher energy-demand. 

\begin{figure}
\centering
\includegraphics[width=0.8\linewidth]{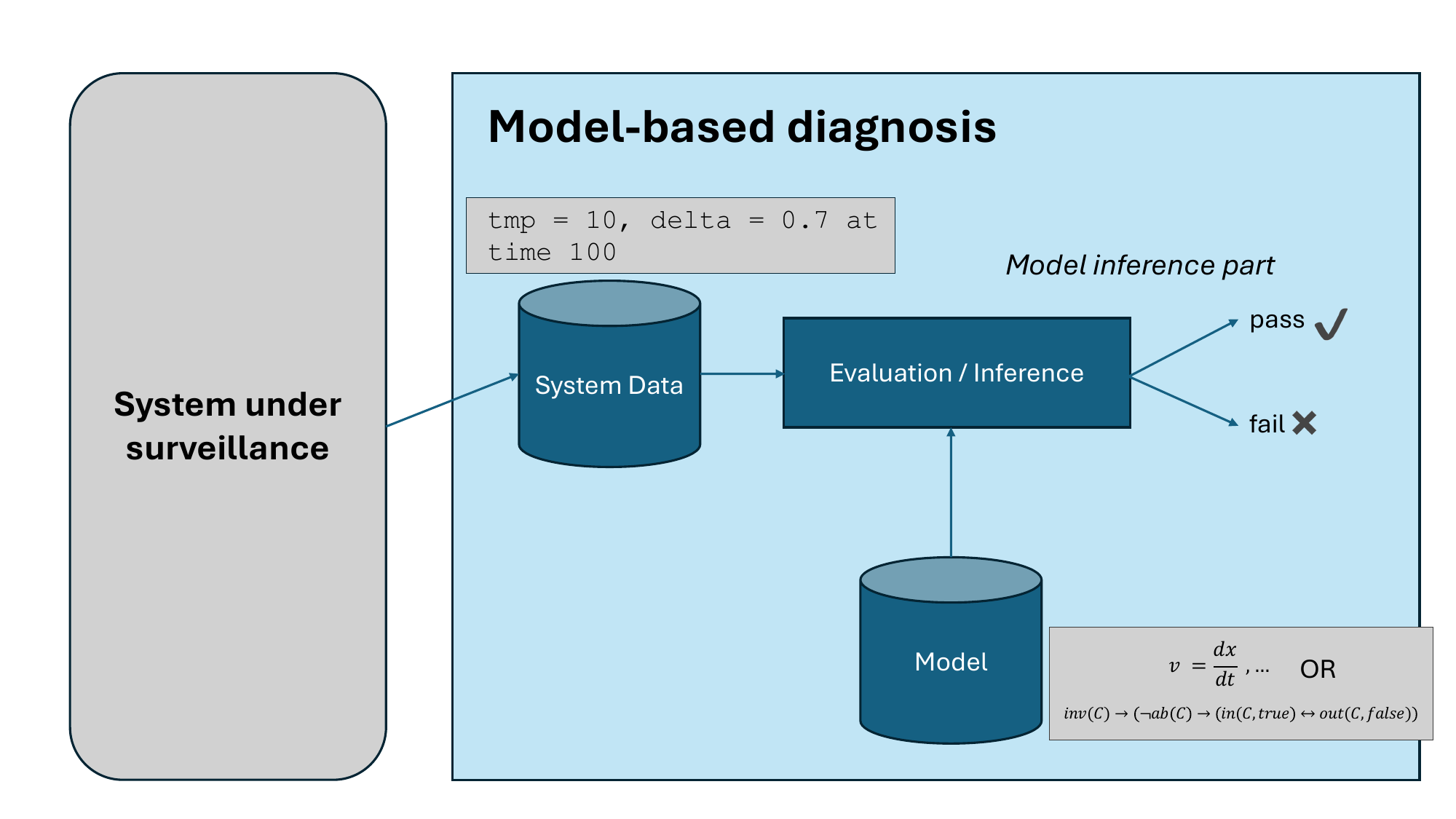}
\caption{Model-based diagnosis architecture}\label{fig:mbdarch}
\end{figure}

Figure~\ref{fig:mbdarch} illustrates the overall architecture of a model-based FDD. Even though its architecture is similar to the architecture of rule-based diagnosis, we see that instead of a rule base a system model is exploited. As a consequence the evaluation component has to be more sophisticated and allow for inference whether the observed values from the system database  contradict the values obtained from the model. For implementing the evaluation/inference component we can follow two principles, (1) the use of logical models and reasoning engines that are either based on consistency~\cite{rei87,kle87,DBLP:conf/ijcai/NicaPQW13} or abduction~\cite{fri90cc,DBLP:journals/ijaeis/Wotawa11,DBLP:journals/apin/Koitz-HristovW20}, or (2) the use of mathematical models, e.g., differential equations. 

{\em Machine learning (ML)} is an essential subfield of AI and has gained a lot of attention, mainly starting in the current century, also in the context of FDD. Many ML methods, such as decision trees~\cite{qui86}, support vector machines (SVM)~\cite{Awad2015}, neural networks~\cite{Haykin2010NeuralNA}, clustering~\cite{JAIN2010651}, regression~\cite{montgomeryEtAl21}, or ensemble methods (e.g., random forests~\cite{598994}) have been used for prediction and classification. In the case of FDD, we are interested in the latter. However, the fundamental idea behind all ML methods is the same: we utilize data to learn a model. 
In particular, we always distinguish two phases in ML, i.e., the training/learning of a model and its use. Both phases come with different ecological footprints as we will discuss in the following section.

\begin{figure}
\centering
\includegraphics[width=0.8\linewidth]{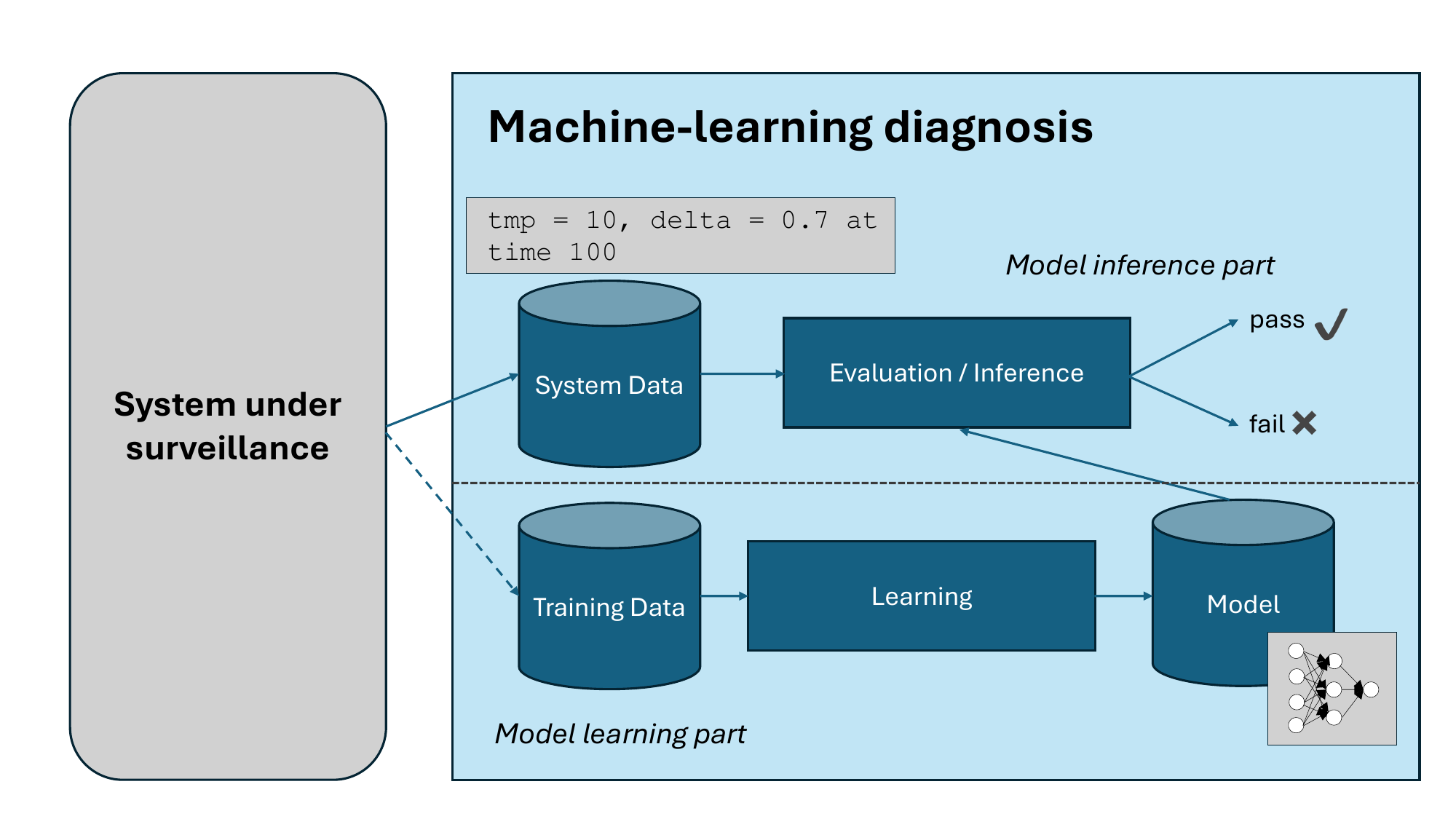}
\caption{Machine learning diagnosis architecture}\label{fig:mlarch}
\end{figure}

Due to distinguishing training (or learning) from inference and reasoning, the software architecture comprises additional components (see Figure~\ref{fig:mlarch}). 
Importantly, we distinguish between training data and system data. The training data is used to learn the model with an algorithm depending on the ML method. The resulting model, similar to model-based diagnosis, is used by an evaluation/inference component, which diagnoses faults based on the model and system data. To work appropriately the system data and the training data should have the same stochastic distribution.

\begin{figure}
\centering
\includegraphics[width=0.8\linewidth]{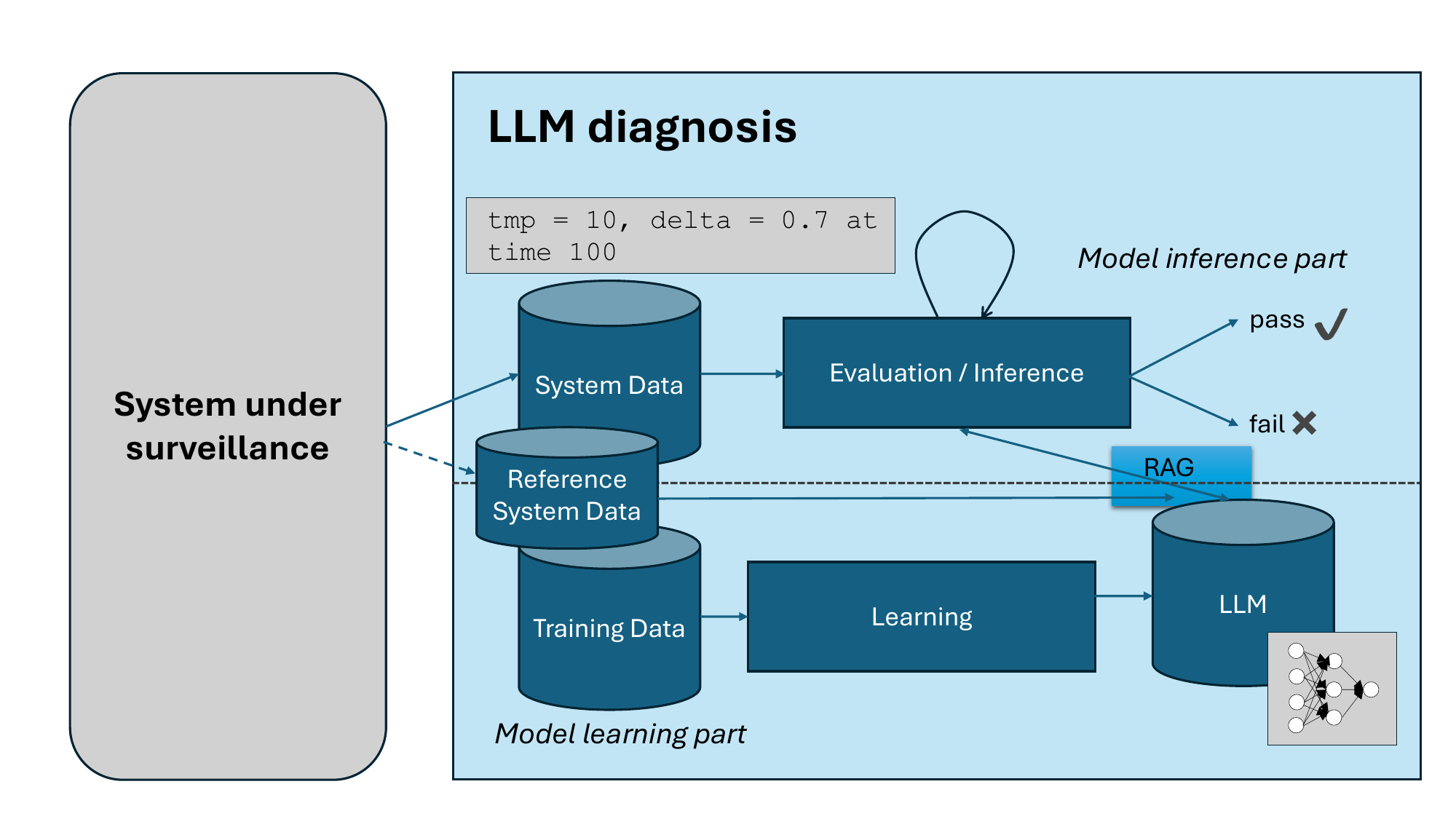}
\caption{LLM-based diagnosis architecture}\label{fig:llmarch}
\end{figure}

Recently, the advances in {\em large language models (LLMs)} have led to an increasing amount of research on LLMs in different application domains, e.g., FDD in BAS~\cite{langerPaper}. LLMs are AI models that predict the next token or sequence of tokens from a given sequence of input tokens. 
Figure~\ref{fig:llmarch} shows the high-level software architecture of an LLM-application for BAS diagnosis. Similar to ordinary ML, there is a component for learning. In practice, the energy-consumption of this component is far higher than for other ML methods, because LLMs are trained on a huge corpus of data and do not focus usually on the system under surveillance. Since LLMs are general-purpose applications, there might be a further training step involved to introduce system specific information---we will later refer to LLMs with additional application-specific training as fine-tuned LLMs. Alternatively, the application might make use of reference system data and RAG in the further evaluation/inference step. Furthermore, inference might be recursively applied until a suitable solution is found, e.g., a classification result can be obtained. Hence, evaluation and inference are also more energy demanding compared to ordinary ML methods. 

\subsection{Evaluation metrics}

Fault detection is a decision problem, i.e., \textit{given the observations is a fault present?} Thus, fault detection can be understood as a simple binary-classification task, i.e., assigning \texttt{nominal} or \texttt{non-nominal} to a set of current observations. 
Based on the degree of missclassifications, we can define the quality of the underlying decision procedure or classification method. For this purpose, we use the following definitions:

\begin{description}
    \item[True positive:] A true positive is an observation from a faulty system that is classified as being faulty.
    \item[True negative:] A true negative is an observation from a correct system that is classified as being correct.
    \item[False positive:] A false positive is a correct observation that is classified as being faulty.
    \item [False negative:] A false negative is a faulty observation that is classified as being correct.
\end{description}

Based on these types of classification, we can count the number of true positives, true negatives, false positives, and false negatives and store this information in $TP$, $TN$, $FP$, and $FN$, respectively. In the rest of this subsection, we use this information, to describe quality metrics of classification methods.

The true positive rate also called sensitivity, is the probability that a test method and in our particular test the fault detection method finds the fault, i.e., the true positive.

$$TPR = \frac{TP}{TP+FN}$$

The false positive rate or type 1 error indicates the probability that the diagnosis method indicates a fault even though there is no fault in the real system.

$$FPR = \frac{FP}{TN+FP}$$

The true negative rate or specificity is the probability that a negative test is correct, i.e., that no fault is present.

$$TNR = \frac{TN}{TN+FP}$$

The false negative rate or type 2 error is the probability that a method misses a fault in a system, i.e., is not able to detect it.

$$FNR = \frac{FN}{TP+FN}$$

It is worth noting that the $FNR$ and the $TPR$ are not independent, i.e., $FNR = 1 - TPR$ holds. Similarly, there is also a dependence between $FPR$ and $TNR$, i.e., $FPR = 1- TNR$.

Alternatively, and originating from the field of information retrieval, there are additional metrics describing the accuracy of an approach, i.e., precision, recall, and the F1 score. Precision is the fraction of the number of true positives divided by the sum of the true positives and false positives. Hence, precision indicates the ratio of the real faults in all faults detected by a diagnosis method.

$$Precision = \frac{TP}{TP+FP}$$

Recall is equivalent to the true positive rate, i.e.:

$$Recall = TPR$$

The F1 score is the harmonic mean of precision and recall.

$$F1 = 2 \cdot \frac{Precision \cdot Recall}{Precision + Recall}$$

Precision, recall, and F1 score are the most commonly used quality metrics reported for classification methods. Unfortunately, from these figures we are not able to compute all the other metrics that are relevant in the ecological model we will present in the next section. For instance, the false positive rate ($FPR$) cannot be computed directly from those other measures. 


\section{Energy-consumption Model}
\label{sec:model}

Answering the question of which diagnosis methodology and straightforwardly its corresponding software architecture to use in order to obtain the most energy-efficient solution requires a detailed analysis of (1) the method's energy-consumption and (2) its effects on the energy-demand of the system to be diagnosed. 
Before developing our energy-consumption model, we need to state its scope.
In our modeling approach, we consider a horizon of one year and assume that the building, its HVAC system and the deployed diagnosis method remain unchanged over this period. In particular, we do not model the aging of components or the associated growth in fault probability of a system's lifetime. Additionally, we represent faults as an aggregate through an annual fraction of wasted energy rather than as a distribution over individual faults. We further restrict our method to consider fault detection, rather than fault identification and repair. 

\subsection{Energy-Consumption Model ($ECM$)}

We start defining the energy-consumption of buildings for cooling and heating $E_{HVAC}$ considering the energy use intensity $EUI$ of a building (i.e., the fault-free required energy per square meter and year), its square meters $A$, and the fraction of the energy-consumption used for heating or cooling $s_{HVAC}$.

\begin{equation}\label{eq:1}
    E_{HVAC} = s_{HVAC} \cdot EUI \cdot A
\end{equation}


Faults often lead to an additional energy-demand that has to be considered. For example, a broken sensor indicating the room has not yet reached the required temperature or a non-optimal setup of parameters in HVAC may cause the system to use more energy for heating. Hence, from a general perspective the additional energy lost is proportional to the optimal energy used for heating and cooling. For computing this wasted energy, we introduce the fraction of energy lost $s_{FAULT}$ in the case of a fault.

\begin{equation} \label{eq:2}
    E_{WASTED} = s_{FAULT} \cdot E_{HVAC}
\end{equation}

It is worth noting that the wasted energy for a fault may be up to 40\%~\cite{bellanco2021Review}. Note further that a more detailed model may consider different types of faults, fault frequency, and fault duration. However, in any case the sum of wasted power over time can be summarized in the annual fraction of energy lost.

The total energy needed for heating and cooling $E_{THVAC}$ of a building in one year without considering diagnosis is stated in the following equation:

\begin{equation} \label{eq:3}
    E_{THVAC} = E_{HVAC} + E_{WASTED}
\end{equation}

After defining the model of energy-consumption of HVAC systems in buildings considering faults, we now focus on the impact of FDD on BAS. In particular, we reduce our view on FDD to fault detection and ignore fault localization, i.e., identifying the root cause of the fault. The reason is that several diagnosis approaches in the context of BAS only consider the detection of an unwanted or unexpected behavior but not on its correction, which often is carried out by maintenance personnel. 
Unfortunately, no approach guarantees a 100\% detection rate. Therefore, we have to consider two errors. First, the case that an FDD approach classifies a value to be wrong but it is not, which raises a wrong alarm. Such a false positive error or type 1 error, causes further investigations but does not necessarily lead to a degraded system. Second, we may not be able to classify a wrong value as indicating a fault. Such a false negative error or type 2 error, does not lead to an alarm and a fault remains. Depending on the classification outcome, we now describe the impact on the energy-consumption.

\begin{description}
\item[True positives (TP):] There is a fault detected, which is a real fault. Hence, when correcting the fault, e.g., by replacing a component or changing a setup, a saving in the energy-consumption, which is proportional to the wasted energy due to faults, is expected. The factor of detected faults is given by the true positive rate $TPR$. In our case, we consider an annual $TPR$, i.e., $TPR_a$, since our time-horizon spans over a single year.  

\begin{equation} \label{eq:4}
    E_{SAVING} = TPR_a \cdot E_{WASTED}
\end{equation}

If all faults can be detected, then there is no energy wasted. Note that in practice, we may still see a wasted energy due to faults, because there is always a delay of repair, which causes additional energy to be wasted. However, compared to the total amount of energy, this remaining wasted energy can be neglected.

\item[True negatives (TN):] There is no fault represented in the observations. Hence, no action has to be taken and thus there is no impact on energy-consumption.

\item[False positives (FP):] False positives lead to unnecessary maintenance activities, which may cause additional energy-consumption. Besides the energy that is needed to perform the repair itself, a false positive may lead to the unnecessary replacement of components still operating as intended. These are, however, secondary energy impacts. 

Overall, we assume a wasted energy for FPs, which is proportional to the false positive rate $FPR$ but only applied to a fraction of the total energy needed for HVAC.  The latter is represented by the factor $s_{FP}$. Again, we consider $FPR$ as an annual rate $FPR_a$.

\begin{equation} \label{eq:5}
    E_{WASTED\_FP} = FPR_a\cdot s_{FP} \cdot E_{HVAC}
\end{equation}


\item[False negatives (FN):] Not identifying all faults causes wasted energy. However, this is already part of the energy-consumption model, since $E_{WASTED}$ already accounts for the additional energy-demand in case of a fault.   

\end{description}

To complete the energy-consumption model considering any diagnosis method, we add the following equation, which defines the total energy-consumption after diagnosis of a BAS.

\begin{equation} \label{eq:6}
    E_{HVAC\_DIAG} = E_{THVAC} - E_{SAVING} + E_{WASTED\_FP} 
\end{equation}



Equations \ref{eq:1} to \ref{eq:6} define the overall energy-consumption model $ECM$. This model already captures important diagnosis-related parameters, i.e., the true positive rate, the false positive rate, and the false negative rate. Other aspects, which are related to the diagnosis methodology are not captured. Table~\ref{tab:model_inputs} depicts all input parameters of the $ECM$ model. 

\begin{table}[t]
\centering
\caption{Input parameters of the energy-consumption model $ECM$ for FDD applied to BAS. $TPR$ and $FPR$ are interpreted as annual averages over the considered year.}
\label{tab:model_inputs}
\begin{tabular}{l p{0.45\linewidth} l l}
\hline
Parameter & Explanation & Unit & Range \\
\hline
$EUI$ & Energy use intensity per year & kWh/(m$^2\cdot$year) & [0,$\infty$) \\
$A$   & Area of building & m$^2$ & [0,$\infty$) \\
$s_{HVAC}$ & Fraction of energy-consumption used for HVAC & 1 & [0,1] \\
$s_{FAULT}$ & Fraction of HVAC energy-consumption in case of faults & 1 & [0,1] \\
$TPR_a$ & Annual true positive rate of diagnosis method & 1 & [0,1] \\
$FPR_a$ & Annual false positive rate of diagnosis method & 1 & [0,1] \\
$s_{FP}$ & Fraction of energy needed to handle FPs & 1 & [0,1] \\
\hline
\end{tabular}

\end{table}

\subsection{Extended Energy-Consumption Model ($EECM$)}

In the following, we extend $ECM$ to consider diagnosis method specific parts, which are influenced by the methods' architectures, discussed in the previous section.
The extended $ECM$, i.e., $EECM$ comprises the information required for model creation and reasoning separately. Let us start with model creation and in particular model learning. 
The energy-consumption depends on the method's energy demands for training $E_{TMETHOD}$, its utility $U$, i.e., the degree to which the model is used exclusively for diagnosis of the BAS, and the number of repetitions necessary per year for keeping the model up to date, i.e., $f_{RET}$. 
\begin{equation}\label{eq:7}
    E_{TRAINING} = E_{TMETHOD} \cdot U \cdot (1 + f_{RET})
\end{equation}
Note that for Equation~\ref{eq:7}, we consider the first year, in which there is one initial training, followed by possibly additional retraining captured by $f_{RET}$. For subsequent years, only $f_{RET}$ would remain. Retraining might be necessary due to data drift~\cite{ZLIOBAITE2015240}. In our $EECM$, we assume that retraining requires the same energy as the original training; this assumption, while reasonable, is not necessarily true the longer the system is in use. For instance, in case additional training data is available, the energy-demand for retraining possibly increases.

The utility $U$ allows us to allocate training energy in proportion to model use, i.e., $U = \frac{N_{INF}}{N_{TOTAL}}$, where $N_{INF}$ is the number of diagnosis inferences per year and $N_{TOTAL}$ encompasses all inferences. Thus, for a model
trained specifically for fault detection, all inferences serve the diagnosis task and $U = 1$, while for a general-purpose model diagnosis may only constitute a part of the total usage.

For inference or reasoning the total energy-consumption depends on the number of iterations per year $N_{INF}$ and the base energy-consumption for one reasoning or inference step $E_{BINF}$. 

\begin{equation}
    E_{INFERENCE} = E_{BINF} \cdot N_{INF} 
\end{equation}

Hence, the total energy-consumption ignoring the energy-consumption of required computing infrastructure etc. is given as follows:

\begin{equation}
    E_{DIAG} = E_{TRAINING} + E_{INFERENCE}
\end{equation}

The total energy-consumption of HVAC diagnosis considering the different methods for the extended $ECM$, which we call $EECM$ is:

\begin{equation}\label{eq:10}
    E_{EXTENDED\_HVAC\_DIAG} = E_{HVAC\_DIAG} + E_{DIAG}
\end{equation}

In Table~\ref{tab:extended_model_inputs}, we summarize the input parameters of the $EECM$. Together with the parameters of the ECM in Table~\ref{tab:model_inputs}, we can determine the annual energy-consumption of a building for cooling and heating. 




\begin{table}[t]
\centering
\caption{Input parameters of the extended energy-consumption model $EECM$ for FDD applied to BAS.}
\label{tab:extended_model_inputs}
\begin{tabular}{l p{0.45\linewidth} l l}
\hline
Parameter & Explanation & Unit & Range \\
\hline
$U$ & Degree to which the model is used exclusively for diagnosis & 1 & (0,1] \\
$f_{RET}$ & Number of additional model retrainings within the considered year & 1 & $\mathbb{N}_0$ \\
$E_{BINF}$ & Energy-consumption of one inference operation& kWh/inference & $[0,\infty)$\\
$N_{INF}$ & Number of inference operations within the considered year& inferences/year & $\mathbb{N}_0$ \\
\hline
\end{tabular}

\end{table}

\section{Simulation Experiments}
\label{sec:experiments}


In our simulation, we consider different architectures and methods for FDD: \em{rule-based} (RB), \em{model-based diagnosis} (MBD), \em{ML}, and \em{LLMs}. For the ML architecture, we review various models at different levels of complexity, i.e., decision trees (DT), random forest (RF), logistic regression (LR), SVM, and multi-layer perceptrons (MLP), i.e., neural networks. For the LLM analysis, we consider LLMs used as-is and fine-tuned LLMs (LLM FT). 

\subsection{Input Data}

For the experiments reported in this paper, we obtain the input values from research on FDD in the building domain. In particular, our starting point was a recent review by Chen et al.~\cite{chen2023review} on data-driven FDD methods for HVAC systems. We considered all comparison papers that perform FDD  on chiller or AHU systems to obtain the rates reported in Table~\ref{tab:diagnosis_tpr_fpr}. To this end, we extracted all usable $TPR$/Recall and $FPR$ values for our considered diagnosis methods from the reviewed studies. Since, however, many studies we examined do not provide the necessary data, i.e., $TPR$/Recall, $FPR$, or a confusion matrix to compute the metrics, we utilized additional chiller- and AHU-specific sources in order to obtain at least a single published value per FDD method. 

Furthermore, the energy model of Section~\ref{sec:model} requires annual values, i.e., $TPR_a$ and $FPR_a$, which describe the behavior of the model over a full year. Even though the reported $TPR$ and $FPR$ from the literature are based on a single experiment or a set of experiments, we use a simplification by assuming that the rates remain stable throughout the year. Thus, for our simulation, we use the range of values we were able to find in the literature for $TPR$/$FPR$ as an approximation for $TPR_a$/$FPR_a$.

\begin{table}[t]
\centering
\caption{Minimum, maximum, and average reported TPR and
FPR for Chiller and AHU FDD methods extracted from the literature reviewed by Chen et al.~\cite{chen2023review}. Summaries of papers used can be found in~\ref{app1}.} 

\label{tab:diagnosis_tpr_fpr}
\begin{tabular}{p{0.25\linewidth} r r r r r r}
\hline
Diagnosis method & \multicolumn{3}{c}{TPR} & \multicolumn{3}{c}{FPR} \\
 & Min & Max & Avg. & Min & Max & Avg.\\
\hline

RB  & 0.11 & 0.39 & 0.25 &0.00 & 0.05 & 0.03$^{*}$\\
(Hybrid)MBD & 0.20$^{a}$ & 1.00$^{a}$& 0.60 & 0.00 & 0.07 & 0.03\\
DT   & 0.17 & 1.00 & 0.61 & $0.00^{b}$  &	0.03$^{c}$  & 0.01 \\
RF   & 0.17 & 1.00 & 0.69 & $0.00$ & $0.00$ & $0.00^{\dagger}$\\
LR   & 0.38 & 0.44& 0.41$^{*}$ & 0.00$^{e}$ & 0.53$^{f}$ & 0.27$^{*}$ \\
SVM  & 0.16 & 1.00 & 0.714 & 0.00 & 0.28 & 0.09\\
MLP  & 0.17 & 1.00 & 0.550 & 0.02 & 0.13 & 0.07$^{*}$\\
LLM$^{d}$ & 0.13 & 0.59 & 0.28 & 0.10 & 0.27 & 0.16\\
LLM FT$^{e,d}$ & 0.98 & 1.00 & 0.99 & 0.00 & 0.02 &0.01 \\ 

\hline
\end{tabular}
\vspace{1mm}
\begin{minipage}{0.98\linewidth}
\footnotesize
Additional sources: ${a}$ Cui and Wang~\cite{CUI2005986}, ${b}$ Padmanabh et al.~\cite{Padmanabh21}, ${c}$ McHugh~\cite{mchugh2019fault}, Zhang et al.~\cite{ZHANG2025124378}${d}$, ${e}$ Langer~\cite{langerThesis}, and ${f}$ Islam et al.~\cite{Islam24}. 

$^{*}$ Due to a limited number of values ($\le 3$), we used the mid-range $(\mathrm{Min}+\mathrm{Max})/2$ to approximate an average value.

$^{\dagger}$ Only one exact numerical FPR value was available, which is hence reported as the midpoint.

\end{minipage}

\end{table}


\begin{table}[t]
\centering

\caption{Energy consumption (in kWh) required for different fault detection methods for a single training/inference run.}

\label{tab:energy}
\begin{tabular}{p{0.12\linewidth} r r r r r r}
\hline
Diagnosis method & \multicolumn{2}{c}{$E_{TMETHOD}$} & $U$ & \multicolumn{2}{c}{$E_{BINF}$} & Ref. \\
 & Min & Max & & Min & Max & \\
\hline
RB$^{a}$ & 0 & 0 & 1 & $1.0\cdot10^{-8}$
& $2.5\cdot10^{-7}$ & \cite{khr27}\\
MBD$^{a}$ & 0 & 0 & 1 & $4.4\cdot10^{-5}$
& $4.6\cdot10^{-4}$ & \cite{koitzhristov_et_al:OASIcs.DX.2025.1}\\
DT & $3.13\cdot10^{-4}$ & $3.13\cdot10^{-4}$ & 1 & $8.60\cdot10^{-5}$  & $8.60\cdot10^{-5}$ & \cite{huertas2023comparative} \\
RF & $2.45\cdot10^{-7}$ & $1.06\cdot10^{-6}$ & 1 & $2.47\cdot10^{-8}$ & $6.35\cdot10^{-8}$ & \cite{Santos24} \\
LR & $3.85\cdot10^{-8}$ & $3.42\cdot10^{-7}$ & 1 & $2.10\cdot10^{-8}$ & $4.41\cdot10^{-8}$ & \cite{Santos24} \\
SVM & $6.04\cdot10^{-8}$ & $3.15\cdot10^{-5}$ & 1 & $3.31\cdot10^{-8}$ & $2.65\cdot10^{-6}$ & \cite{Santos24} \\
MLP & $2.69\cdot10^{-3}$ & $3.24\cdot10^{-2}$ & 1 & $4.50\cdot10^{-5}$ & $1.15\cdot10^{-4}$ & \cite{huertas2023comparative} \\
LLM & $1.5\cdot10^3$ & $62.32\cdot10^6$ & $1.1\cdot 10^{-7}$ & $2\cdot10^{-6}$ & $2.9\cdot10^{-3}$ & \cite{JI2026116159}$^b$ \\
LLM FT & $1.65\cdot10^3$ & $68.55\cdot10^6$ & $1.1\cdot 10^{-7}$ & $2\cdot10^{-6}$ & $2.9\cdot10^{-3}$ & \cite{JI2026116159}$^c$ \\
\hline
\end{tabular}

\vspace{1mm}
\begin{minipage}{0.98\linewidth}
\footnotesize
$^{a}$ Estimated energy-consumption.\\
$^b$ See Table 3 and 5 of the mentioned paper. Inference energy-consumption also considers more sophisticated queries to capture agentic systems requiring more than one LLM query for providing an outcome. \\
$^c$ Like for ordinary LLMs. For fine tuning 10\% more energy for training assumed. 
\end{minipage}

\end{table}

Besides the information on the prediction accuracy of the considered architectures and methods, our extended energy model $EECM$ relies on information on the energy-consumption of training the model (i.e., $E_{TMETHOD}$), the utility factor $U$, and the energy consumed by performing a single inference step (i.e., $E_{BINF}$). Table~\ref{tab:energy} depicts those figures based on prior studies not necessarily in the building domain. 
As one can observe, for RB and MBD we assign a training energy of zero, since neither involves a learning stage. In both cases, the knowledge is derived by domain experts rather than from data, thus, no computation is performed before deployment. 

In addition, the energy-consumption of inference for RB and MBD was estimated from the runtimes reported in the cited studies. Runtime is not a direct measure of environmental impact; however, studies~\cite{Bruce,Zhou18,Weber23} indicate that it can serve as a proxy for energy consumption. Since all experiments~\cite{khr27,koitzhristov_et_al:OASIcs.DX.2025.1} were conducted on the same machine\footnote{Mac Pro equipped with a 12-core Intel Xeon E5 processor at 2.7\,GHz and 64\,GB RAM} we estimate the energy-consumption from the reported runtime as $ E_{\mathrm{est}} = P_{\mathrm{est}}\frac{t}{3600}, $ where $t$ denotes the runtime in seconds and $P_{\mathrm{est}}$ the estimated average computational power in kW. An average power draw of $P_{\mathrm{est}}=0.12$\,kW was obtained from CodeCarbon~\cite{codecarbon} measurements performed on the same machine with similar experiments. 

Considering the utility $U$, we assign a value of one to all methods except LLMs. The reason is that in the other cases, whether those are rules or ML models, they are created for a specific purpose. Hence, those systems cannot be used for an arbitrary task, whereas LLMs can be used for different applications without, in theory, requiring additional training. Hence, for LLMs we approximate $U$ based on the widely used deployed conversational model ChatGPT.  Approximately $2.5 \cdot 10^{9}$ prompts are processed per day~\cite{silberling2025prompts}. Thus, there are  around $N_{TOTAL} \approx 9.1 \cdot 10^{11}$ queries per year. Assuming $N_{INF} = 10^{5}$ diagnostic queries annually yields a utility of $\approx 1.1\cdot10^{-7}$. While one can argue that the model would have been trained regardless of the diagnosis applications using it, we consider a proportional training. In the case of the fine-tuned LLM, we add an additional 10\% to account for the fine-tuning.


\subsection{Simulation Results}


To carry out the experiments, we implemented both models and also all experimental setups in R~\cite{R} version 4.3.2. The source code is available at Zenodo~\cite{wotawa_2026_22105577}. We used a MacBook Pro with an Apple M4 Pro processor and 48 GB main memory and macOS Tahoe 26.6.1.

\paragraph{First Experiment}
\begin{sloppypar}
    In the initial experiment, we estimated the energy-consumption for all methods assuming the energy needed for a mid-sized office building to be $250~kWh/m^2$ and a varying total number of square meters ranging from 1,000 to 10,000$m^2$. We further assume that $s_{hvac}$ is uniformly distributed between 0.1 and 0.8, $s_{fault}$ between 0.1 and 0.4, and $s_{fp}$ between 0.01 and 0.1. For the other parameters we used the minimum and maximum values stated in Tables~\ref{tab:diagnosis_tpr_fpr} and \ref{tab:energy} and again assume a random uniform distribution. Furthermore, we assume that there is only one training needed and that there is a measurement every 10 minutes. For this and all the other experiments, we created 1,000,000 random instances using the mentioned parameters and their ranges.
\end{sloppypar}

\begin{table}
\scriptsize
\centering
\caption{The energy-consumption calculated for the different FDD methods considering different building areas $A$ using $ECM$ and $EECM$ together with their computed potential savings in percent.} 
\label{tab:results1}
\begin{tabular}{rrlrrrrr}
  \hline
 & $A$ & Method & $e_{thvac}$ & $e_{hvac\_diag}^{ECM}$ & $e_{hvac\_diag}^{EECM}$ & $saving^{ECM}$ & $saving^{EECM}$ \\ 
  \hline
1 & 1000.00 & RB & 140584.96 & 133715.29 & 133715.30 & 4.89 & 4.89 \\ 
  2 & 1000.00 & MBD & 140584.96 & 123932.96 & 123946.20 & 11.84 & 11.84 \\ 
  3 & 1000.00 & DT & 140584.96 & 124226.09 & 124230.61 & 11.64 & 11.63 \\ 
  4 & 1000.00 & RF & 140584.96 & 124133.26 & 124133.26 & 11.70 & 11.70 \\ 
  5 & 1000.00 & LR & 140584.96 & 130696.54 & 130696.55 & 7.03 & 7.03 \\ 
  6 & 1000.00 & SVM & 140584.96 & 125138.82 & 125138.89 & 10.99 & 10.99 \\ 
  7 & 1000.00 & MLP & 140584.96 & 124612.48 & 124616.70 & 11.36 & 11.36 \\ 
  8 & 1000.00 & LLM & 140584.96 & 131598.32 & 131678.10 & 6.39 & 6.34 \\ 
  9 & 1000.00 & LLM FT & 140584.96 & 112808.76 & 112888.80 & 19.76 & 19.70 \\ \hline 
  10 & 2500.00 & RB & 351462.39 & 334271.80 & 334271.81 & 4.89 & 4.89 \\ 
  11 & 2500.00 & MBD & 351462.39 & 309833.46 & 309846.72 & 11.84 & 11.84 \\ 
  12 & 2500.00 & DT & 351462.39 & 310565.62 & 310570.14 & 11.64 & 11.63 \\ 
  13 & 2500.00 & RF & 351462.39 & 310360.13 & 310360.13 & 11.69 & 11.69 \\ 
  14 & 2500.00 & LR & 351462.39 & 326738.53 & 326738.54 & 7.03 & 7.03 \\ 
  15 & 2500.00 & SVM & 351462.39 & 312898.31 & 312898.38 & 10.97 & 10.97 \\ 
  16 & 2500.00 & MLP & 351462.39 & 311502.08 & 311506.30 & 11.37 & 11.37 \\ 
  17 & 2500.00 & LLM & 351462.39 & 329019.95 & 329099.71 & 6.39 & 6.36 \\ 
  18 & 2500.00 & LLM FT & 351462.39 & 282021.64 & 282101.68 & 19.76 & 19.73 \\  \hline 
  19 & 5000.00 & RB & 702924.78 & 668566.56 & 668566.57 & 4.89 & 4.89 \\ 
  20 & 5000.00 & MBD & 702924.78 & 619599.53 & 619612.78 & 11.85 & 11.85 \\ 
  21 & 5000.00 & DT & 702924.78 & 621191.87 & 621196.39 & 11.63 & 11.63 \\ 
  22 & 5000.00 & RF & 702924.78 & 620689.99 & 620689.99 & 11.70 & 11.70 \\ 
  23 & 5000.00 & LR & 702924.78 & 653487.94 & 653487.95 & 7.03 & 7.03 \\ 
  24 & 5000.00 & SVM & 702924.78 & 625690.03 & 625690.10 & 10.99 & 10.99 \\ 
  25 & 5000.00 & MLP & 702924.78 & 623070.25 & 623074.47 & 11.36 & 11.36 \\ 
  26 & 5000.00 & LLM & 702924.78 & 658031.17 & 658110.87 & 6.39 & 6.38 \\ 
  27 & 5000.00 & LLM FT & 702924.78 & 564044.76 & 564124.84 & 19.76 & 19.75 \\ \hline 
  28 & 7500.00 & RB & 1054387.17 & 1002832.39 & 1002832.40 & 4.89 & 4.89 \\ 
  29 & 7500.00 & MBD & 1054387.17 & 929476.58 & 929489.82 & 11.85 & 11.85 \\ 
  30 & 7500.00 & DT & 1054387.17 & 931730.52 & 931735.04 & 11.63 & 11.63 \\ 
  31 & 7500.00 & RF & 1054387.17 & 931006.16 & 931006.17 & 11.70 & 11.70 \\ 
  32 & 7500.00 & LR & 1054387.17 & 980218.78 & 980218.79 & 7.03 & 7.03 \\ 
  33 & 7500.00 & SVM & 1054387.17 & 938488.90 & 938488.97 & 10.99 & 10.99 \\ 
  34 & 7500.00 & MLP & 1054387.17 & 934476.95 & 934481.18 & 11.37 & 11.37 \\ 
  35 & 7500.00 & LLM & 1054387.17 & 987055.75 & 987135.41 & 6.39 & 6.38 \\ 
  36 & 7500.00 & LLM FT & 1054387.17 & 846068.73 & 846148.77 & 19.76 & 19.75 \\ \hline
  37 & 10000.00 & RB & 1405849.55 & 1337118.70 & 1337118.71 & 4.89 & 4.89 \\ 
  38 & 10000.00 & MBD & 1405849.55 & 1239392.96 & 1239406.21 & 11.84 & 11.84 \\ 
  39 & 10000.00 & DT & 1405849.55 & 1242282.22 & 1242286.74 & 11.63 & 11.63 \\ 
  40 & 10000.00 & RF & 1405849.55 & 1241465.55 & 1241465.55 & 11.69 & 11.69 \\ 
  41 & 10000.00 & LR & 1405849.55 & 1306978.81 & 1306978.81 & 7.03 & 7.03 \\ 
  42 & 10000.00 & SVM & 1405849.55 & 1251363.91 & 1251363.98 & 10.99 & 10.99 \\ 
  43 & 10000.00 & MLP & 1405849.55 & 1246055.11 & 1246059.34 & 11.37 & 11.37 \\ 
  44 & 10000.00 & LLM & 1405849.55 & 1316085.38 & 1316165.07 & 6.39 & 6.38 \\ 
  45 & 10000.00 & LLM FT & 1405849.55 & 1128086.80 & 1128166.75 & 19.76 & 19.75 \\ 
   \hline
\end{tabular}
\end{table}

Table~\ref{tab:results1} shows the obtained estimates of the mean energy-consumption of the overall building's HVAC with and without diagnosis using $ECM$ and $EECM$. 
Obviously, the relative energy savings for the $ECM$ do not change for the different building areas, because all terms scale proportionally to the area. Thus, differences between the accuracy of the underlying FDD methods only depend on their diagnostic performance, i.e., $TPR$ and $FPR$. For the $ECM$ we see the following ordering of methods when using the energy savings: LLM FT with almost 20\% followed by MBD, RF, DT, MLP, and SVM all with approximately than 11\% saving, and finally LR, LLM, and RB. The overall savings range from about 5 to 20\% depending on the methods. 

For $EECM$ there is also no difference between the savings for different areas of the building. Moreover, both models yield almost identical savings. This indicates that the energy-consumption of training and inference for the assumed values in the first experiment can be neglected. 

\begin{figure}[t]
\begin{center}
\begin{tikzpicture}
\begin{axis}[
    title=Savings as function of the number of trainings,
    xmode=normal,
    ymode=normal,
    xlabel={number of additional trainings},
    ylabel={savings in \%},
    legend style={
cells={anchor=east},
legend pos=outer north east,
},
]

 \addplot [red,thick,dotted,name path=A,domain=0:3] table[col sep=comma, x="fret",y="RB"] {data/test2.csv};
 \addlegendentry{RB}
 \addplot [green,thick,name path=D,domain=0:3] table[col sep=comma, x="fret",y="MBD"] {data/test2.csv};
 \addlegendentry{MBD}
 \addplot [blue,thick, dashed, name path=A,domain=0:3] table[col sep=comma, x="fret",y="DT"] {data/test2.csv};
 \addlegendentry{DT}
 \addplot [cyan,thick,dashdotted,name path=D,domain=0:3] table[col sep=comma, x="fret",y="RF"] {data/test2.csv};
 \addlegendentry{RF}
  \addplot [magenta,thick,loosely dotted,name path=A,domain=0:3] table[col sep=comma, x="fret",y="LR"] {data/test2.csv};
  \addlegendentry{LR}
 \addplot [violet,thick,loosely dashed,name path=D,domain=0:3] table[col sep=comma, x="fret",y="SVM"] {data/test2.csv};
 \addlegendentry{SVM}
 \addplot [black,thick,name path=A,domain=0:3] table[col sep=comma, x="fret",y="MLP"] {data/test2.csv};
 \addlegendentry{MLP}
 \addplot [orange,thick,dashdotdotted,name path=D,domain=0:3] table[col sep=comma, x="fret",y="LLM"] {data/test2.csv};
 \addlegendentry{LLM}
  \addplot [brown,thick,loosely dashdotted, name path=D,domain=0:3] table[col sep=comma, x="fret",y="LLM.FT"] {data/test2.csv};
  \addlegendentry{LLM FT}
  
\end{axis}
\end{tikzpicture}
\end{center}
\vspace{-1em}
\caption{Results of the second experiment.}
\label{fig:test2}
\end{figure}
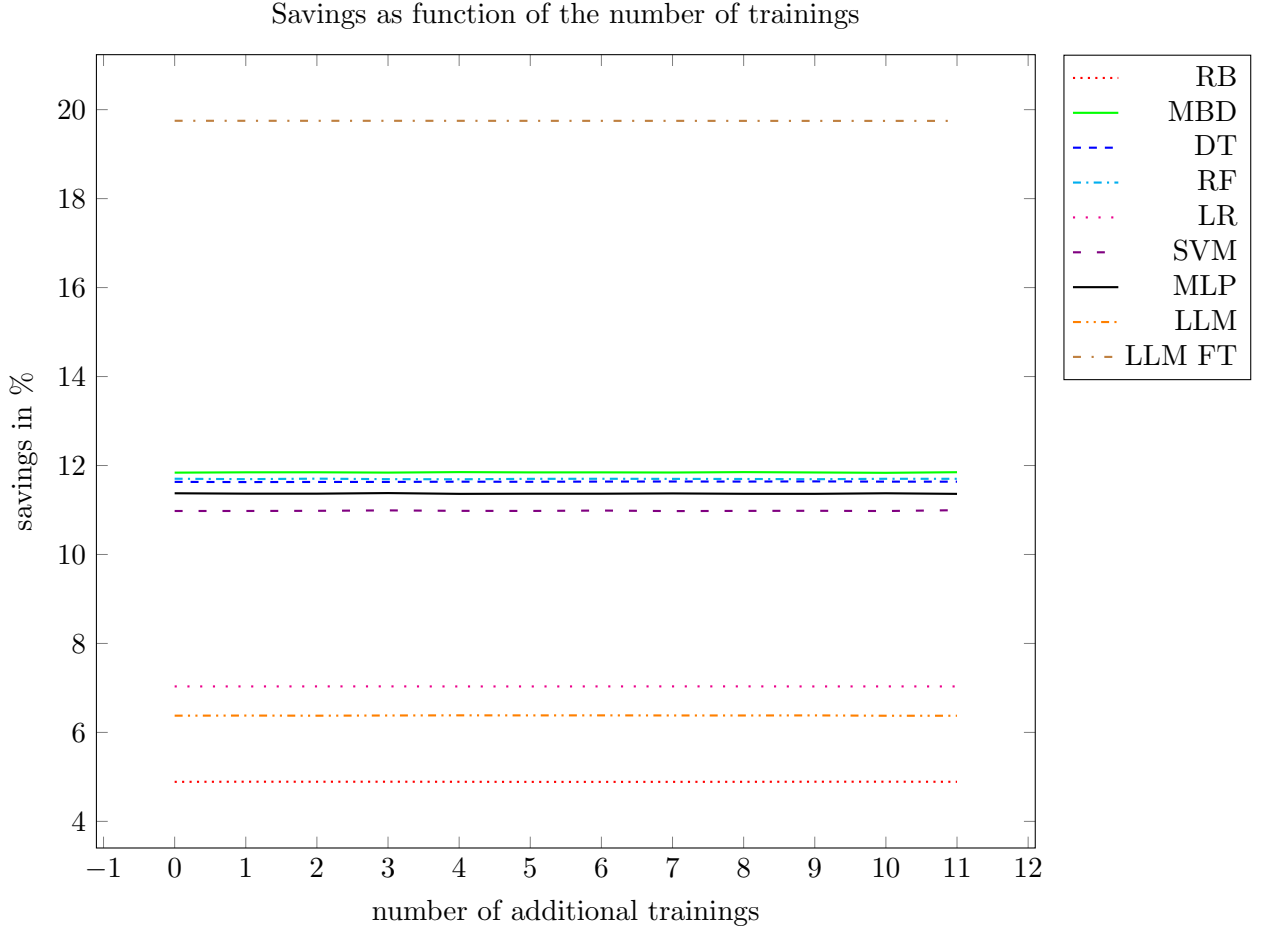

\paragraph{Second Experiment}
In the second experiment, we focused on the dependency of the number of trainings over time on the impact on diagnosis. Hence, we only consider the $EECM$. We set the overall building area $A$ to 10,000. We range the number of trainings from 1 (which means 0 repetitions) to 12 (11 training repetitions), i.e., every month we have a training. All other parameters remain the same as for the initial experiment. Figure~\ref{fig:test2} depicts the outcome of the second experiment. We see no impact on any method when varying the number of trainings for the building sector. 

\paragraph{Third Experiment}
The third experiment is similar to the second. Instead of varying the number of trainings, we vary the number of inferences from 1 minute to 60 minutes. The building area remains 10,000, and we assume only one training. The other parameters remain the same. The selected inference intervals are based on experience with building monitoring and sensor-data collection.

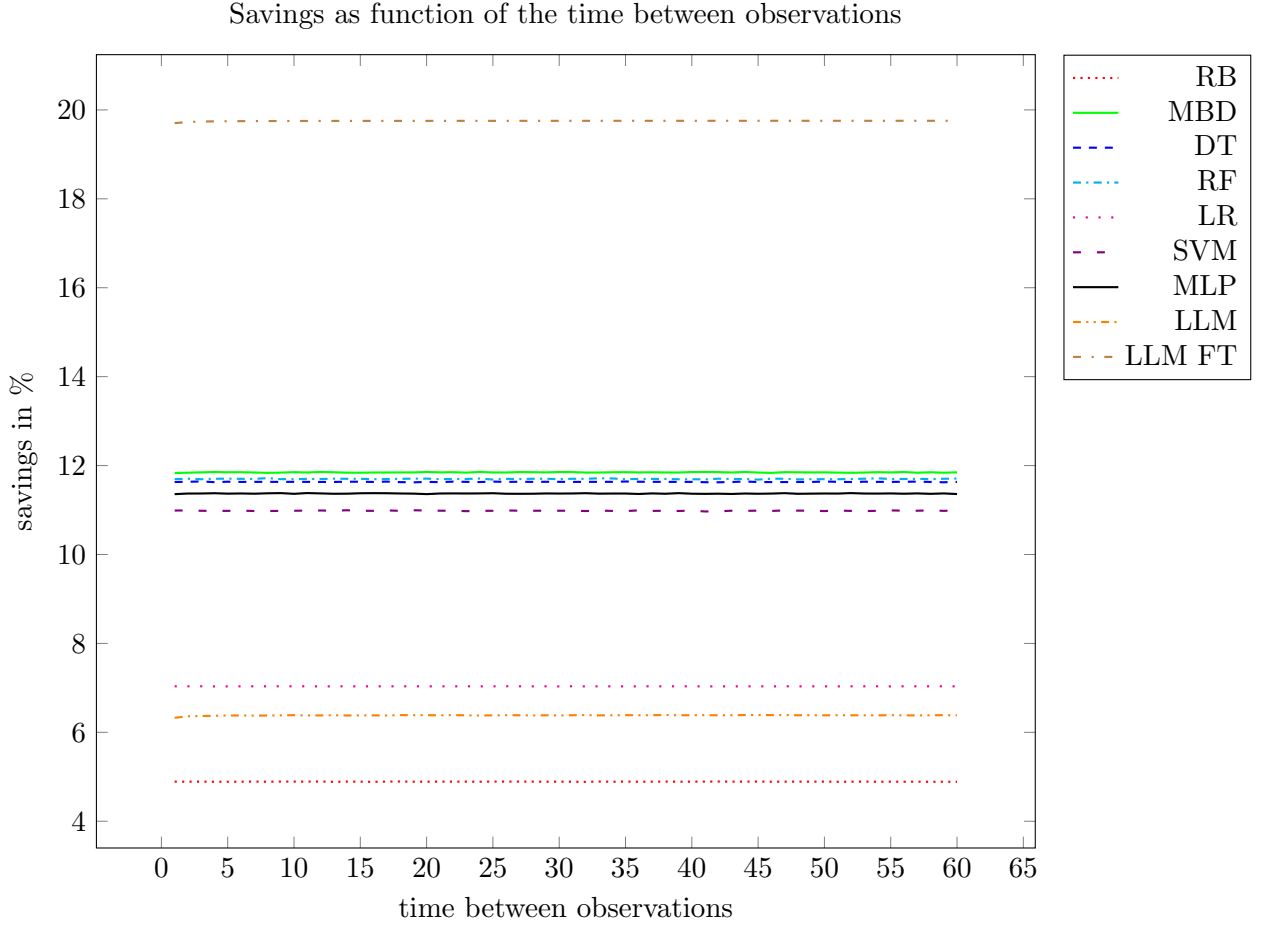
\begin{figure}[t]
\begin{center}
\begin{tikzpicture}
\begin{axis}[
    title=Savings as function of the time between observations,
    xmode=normal,
    ymode=normal,
    xlabel={time between observations},
    ylabel={savings in \%},
    legend style={
cells={anchor=east},
legend pos=outer north east,
},
]

 \addplot [red,thick,dotted,name path=A,domain=0:3] table[col sep=comma, x="ts",y="RB"] {data/test3.csv};
 \addlegendentry{RB}
 \addplot [green,thick,name path=D,domain=0:3] table[col sep=comma, x="ts",y="MBD"] {data/test3.csv};
 \addlegendentry{MBD}
 \addplot [blue,thick, dashed, name path=A,domain=0:3] table[col sep=comma, x="ts",y="DT"] {data/test3.csv};
 \addlegendentry{DT}
 \addplot [cyan,thick,dashdotted,name path=D,domain=0:3] table[col sep=comma, x="ts",y="RF"] {data/test3.csv};
 \addlegendentry{RF}
\addplot [magenta,thick,loosely dotted,name path=A,domain=0:3] table[col sep=comma, x="ts",y="LR"] {data/test3.csv};
  \addlegendentry{LR}
 \addplot [violet,thick,loosely dashed,name path=D,domain=0:3] table[col sep=comma, x="ts",y="SVM"] {data/test3.csv};
 \addlegendentry{SVM}
 \addplot [black,thick,name path=A,domain=0:3] table[col sep=comma, x="ts",y="MLP"] {data/test3.csv};
 \addlegendentry{MLP}
 \addplot [orange,thick,dashdotdotted,name path=D,domain=0:3] table[col sep=comma, x="ts",y="LLM"] {data/test3.csv};
 \addlegendentry{LLM}
  \addplot [brown,thick,loosely dashdotted, name path=D,domain=0:3] table[col sep=comma, x="ts",y="LLM.FT"] {data/test3.csv};
  \addlegendentry{LLM FT}
  
\end{axis}
\end{tikzpicture}
\end{center}
\vspace{-1em}
\caption{Results of the third experiment.}
\label{fig:test3}
\end{figure}

Figure~\ref{fig:test3} shows the results of the third experiment. Again, we do not see any impact. Hence, for the building sector only the $TPR$ and $FPR$ matter for selecting the best diagnosis method and their corresponding software architecture.

\paragraph{Fourth Experiment}
he fourth experiment examines whether using mean $TPR$ and $FPR$ values instead of maximum values affects the estimated savings. As in the previous experiments, we still consider office buildings and use a uniform distribution for the following parameters: area ranging from 1,000 to 10,000, energy consumption per year from 200-300kWh/m$^2$, observation time from 1 to 60 minutes, and the number of training steps in total ranging from 1 to 12. The other parameters remain the same as in the first experiment.

\begin{table}[ht]
\caption{The results of the 4th experiment.} 
\label{tab:test4}
\centering
\begin{tabular}{rlrrrrr}
  \hline
 & Method & $e_{thvac}$ & $e_{hvac\_diag}^{ECM}$ & $e_{hvac\_diag}^{EECM}$ & $saving^{ECM}$ & $saving^{EECM}$ \\ 
  \hline
1 & RB & 774327.98 & 736628.85 & 736628.86 & 4.87 & 4.87 \\ 
  2 & MBD & 774327.98 & 682419.45 & 682428.60 & 11.87 & 11.87 \\ 
  3 & DT & 774327.98 & 680189.37 & 680192.49 & 12.16 & 12.16 \\ 
  4 & RF & 774327.98 & 667458.02 & 667458.03 & 13.80 & 13.80 \\ 
  5 & LR & 774327.98 & 720022.36 & 720022.36 & 7.01 & 7.01 \\ 
  6 & SVM & 774327.98 & 666806.41 & 666806.46 & 13.89 & 13.89 \\ 
  7 & MLP & 774327.98 & 691526.14 & 691529.17 & 10.69 & 10.69 \\ 
  8 & LLM & 774327.98 & 736410.43 & 736485.47 & 4.90 & 4.89 \\ 
  9 & LLM FT & 774327.98 & 621333.45 & 621410.80 & 19.76 & 19.75 \\ 
   \hline
\end{tabular}
\end{table}

Table~\ref{tab:test4} depicts the outcome. There is no difference in the overall ranking. It is worth noting that there is a minor decrease in savings for the LLM and LLM FT due to the high energy-consumption required for training. This drop is, however, minor. For the other methods the computed savings are the same regardless of using $ECM$ or $EECM$.



\paragraph{Fifth Experiment}
In the experiment, we assume a single-family house considering uniform distributions of the following parameters: usable floor area from 100-200m$^2$, total energy-consumption of 50-100kWh/m$^2$, time between observations from 1 to 60 minutes, and number of trainings ranging from 1 to 12 in total. We further adapted $s_{HVAC}$ to range from 0.6 to 0.8 covering the fact that most of the energy in households is used for heating and cooling. The remaining parameters are the same as for the first experiment.

\begin{table}[ht]
\caption{The results of the 5th experiment for single-family homes.} 
\label{tab:test5}
\centering
\begin{tabular}{rlrrrrr}
  \hline
 & Method & $e_{thvac}$ & $e_{hvac\_diag}^{ECM}$ & $e_{hvac\_diag}^{EECM}$ & $saving^{ECM}$ & $saving^{EECM}$ \\ 
  \hline
1 & RB & 8531.38 & 8115.97 & 8115.97 & 4.87 & 4.87 \\ 
  2 & MBD & 8531.38 & 7518.62 & 7527.77 & 11.87 & 11.76 \\ 
  3 & DT & 8531.38 & 7494.05 & 7497.18 & 12.16 & 12.12 \\ 
  4 & RF & 8531.38 & 7353.76 & 7353.76 & 13.80 & 13.80 \\ 
  5 & LR & 8531.38 & 7932.98 & 7932.98 & 7.01 & 7.01 \\ 
  6 & SVM & 8531.38 & 7346.58 & 7346.63 & 13.89 & 13.89 \\ 
  7 & MLP & 8531.38 & 7618.97 & 7621.99 & 10.69 & 10.66 \\ 
  8 & LLM & 8531.38 & 8113.56 & 8188.57 & 4.90 & 4.02 \\ 
  9 & LLM FT & 8531.38 & 6845.51 & 6922.69 & 19.76 & 18.86 \\ 
   \hline
\end{tabular}
\end{table}

Table~\ref{tab:test5} shows the obtained results. Due to the smaller amount of total energy-consumption, we now see a difference between the savings considering $ECM$ and $EECM$. For this class of buildings, considering the energy demand of inference and trainings makes a difference, but does not change the overall picture significantly. Again LLM FT performs best in terms of savings, followed by SVM and RF (which are close together). Afterward, we have DT, MBD, MLP, RB, and finally LLM without fine tuning. Due to the low $TPR$ of LLM and the energy needs for training and inference, it is ranked last.

\paragraph{Sixth Experiment}
In the sixth experiment we used the same parameter values as in the fifth experiment with one exception. We do not randomly distribute the time between observations, but use it as a parameter to illustrate the impact of the number of inference steps on the overall energy-consumption and savings.

\begin{figure}[t]
\begin{center}
\begin{tikzpicture}
\begin{axis}[
    title=Savings as function of the time between observations,
    xmode=log,
    ymode=normal,
    xlabel={time between observations},
    ylabel={savings in \%},
    legend style={
cells={anchor=east},
legend pos=outer north east,
},
]

 \addplot [red,thick,dotted,name path=A,domain=0:3] table[col sep=comma, x="ts",y="RB"] {data/test6.csv};
 \addlegendentry{RB}
 \addplot [green,thick,name path=D,domain=0:3] table[col sep=comma, x="ts",y="MBD"] {data/test6.csv};
 \addlegendentry{MBD}
 \addplot [blue,thick, dashed, name path=A,domain=0:3] table[col sep=comma, x="ts",y="DT"] {data/test6.csv};
 \addlegendentry{DT}
 \addplot [cyan,thick,dashdotted,name path=D,domain=0:3] table[col sep=comma, x="ts",y="RF"] {data/test6.csv};
 \addlegendentry{RF}
\addplot [magenta,thick,loosely dotted,name path=A,domain=0:3] table[col sep=comma, x="ts",y="LR"] {data/test6.csv};
  \addlegendentry{LR}
 \addplot [violet,thick,loosely dashed,name path=D,domain=0:3] table[col sep=comma, x="ts",y="SVM"] {data/test6.csv};
 \addlegendentry{SVM}
 \addplot [black,thick,name path=A,domain=0:3] table[col sep=comma, x="ts",y="MLP"] {data/test6.csv};
 \addlegendentry{MLP}
 \addplot [orange,thick,dashdotdotted,name path=D,domain=0:3] table[col sep=comma, x="ts",y="LLM"] {data/test6.csv};
 \addlegendentry{LLM}
  \addplot [brown,thick,loosely dashdotted, name path=D,domain=0:3] table[col sep=comma, x="ts",y="LLM.FT"] {data/test6.csv};
  \addlegendentry{LLM FT}
  
\end{axis}
\end{tikzpicture}
\end{center}
\vspace{-1em}
\caption{Results of the sixth experiment.}
\label{fig:test6}
\end{figure}
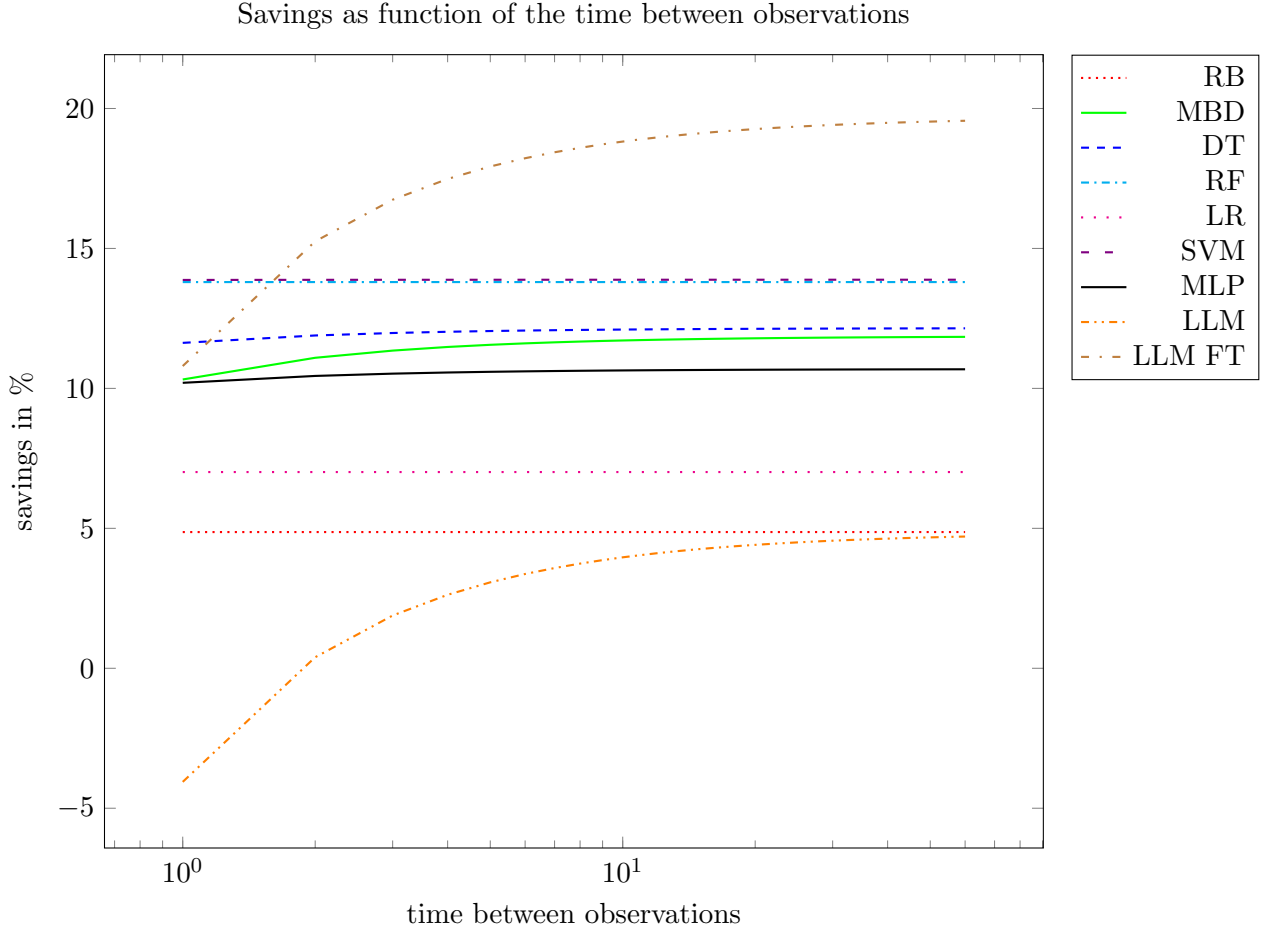

\begin{sloppypar}

Figure~\ref{fig:test6} depicts the outcome and shows the dependency of energy-consumption on the time between observations for single-family houses. Note that the x-axis is log-scaled. From this figure we see that if we apply inference every minute, approaches such as RF and SVM are superior. If this time is increased to 2 minutes, LLM FT takes over. Hence, fine-tuned LLMs are still the method of choice when considering more realistic inference frequencies of about 5 to 10 minutes for buildings. However, for other application areas where there is a higher demand for detecting faults, other FDD approaches might provide more total energy savings.
\end{sloppypar}


\section{Limitations}\label{sec:threats}

Our study combines a theoretical energy-consumption model with simulation experiments using values from prior literature. The validity of the results and conclusions drawn therefore depends on the model's underlying assumptions and simplifications, as well as the quality of the input data.

\subsection{Model assumptions}
As with any theoretical model, our proposed $ECM$ and $EECM$ rely on several assumptions and simplifications that limit how accurately they represent a concrete building and its HVAC system.
First, we consider a one-year time horizon rather than the entire building life cycle. During this period, we assume the building, its HVAC system, and the FDD methods used remain unchanged. Consequently, our models do not capture long-term changes, such as building modifications, replacement or degradation of HVAC components, or updates to the FDD system. We also do not model component aging and the related increase in fault probability. Given these simplifications, the energy-consumption models should be interpreted as approximations rather than precise representations of a building's behavior over its full lifetime. 

Second, we combine fault effects into a single parameter. Instead of modeling individual fault types with different severities, occurrence frequencies, and durations, we approximate the effects of all faults as a fraction of the annual HVAC energy loss. This simplification allows us to compare the different diagnosis architectures and methods without requiring a detailed fault model for each specific HVAC system. Consequently, our models do not account for the distinct ecological impacts of individual faults.

Third, our models do not consider fault identification, localization, and repair. We assume that once a fault is detected, it is immediately corrected and the associated energy savings take effect. In practice, however, detecting a fault is often not sufficient. In order to restore the nominal system behavior the faulty component has to be identified. Furthermore, there is typically a delay between fault detection and the corresponding repair action. During this interval, the fault persists and continues to waste energy. Our models do not factor in the additional energy and resources that may be attributed to the repair and replacement activity, e.g., recommissioning. We assume these effects on the operational energy are negligible compared to the annual HVAC energy-consumption. 

Finally, as noted above, our models capture building scale through area and energy use intensity but do not explicitly represent the complexity of the HVAC system. Two buildings with a similar area and energy demand may differ in the number of HVAC components (i.e., the possible subsystems that may experience faults) and the effort required to diagnose and maintain the system. Our models do not directly account for this.


\subsection{Validity of input data and parameter estimates}
Our simulation experiments are based on values obtained from prior literature on the diagnostic accuracy of FDD methods and the ecological impact of various architectures and methods. The reliability of our results therefore depends on the quality, comparability, and representativeness of the underlying input data. While we rely on prior published research for our data collection, it was not conducted as an exhaustive systematic review; consequently, other relevant studies may contain values outside the ranges reported. 

\paragraph{Diagnostic performance data} 
Our starting point was a recent review by Chen et al.~\cite{chen2023review}, which analyzes research on FDD diagnosis techniques for HVAC systems. Thus, the studies selected from the review differ with regard to the HVAC systems under investigation, data sets, fault types, operating conditions, experimental setups, and the diagnosis tasks considered (i.e., fault detection or identification). Consequently, the values we obtained are not directly comparable, but they reflect the heterogeneity across HVAC FDD.

The availability of suitable performance data has been an issue. Many studies we examined do not report the required metrics (i.e., $TPR$, $FPR$, or confusion matrices from which these metrics can be derived), nor do they provide reproducibility packages containing the results. Therefore, we included additional sources where necessary. Yet, for methods where only a few values have been compiled, the resulting estimates carry greater uncertainty. 

Another simplification concerns the temporal interpretation of the reported performance values. The literature values are rooted in individual experiments or sets of experiments rather than observations over an entire year. Yet our models require the annual true positive and false positive rates. We therefore assume the reported values are suitable approximations for $TPR_a$ and $FPR_a$, although in practice, diagnostic performance over a year may vary, e.g., due to data drift.
We partially address these concerns by documenting the data extraction and relying on minimum, maximum, and average values rather than individual values reported in the literature. 

\paragraph{Computational energy data}
As with the diagnostic performance data, we obtained the computational energy values from prior literature. However, we did not limit our analysis to HVAC applications, since application-specific energy measurements are sparse in the building domain. Consequently, the reported values stem from different application domains, hardware platforms, software frameworks, model configurations, and experimental setups. We acknowledge that all of these factors can substantially influence an FDD method's energy-consumption. However, we retain the values and use their ranges to capture the variability.

In addition, suitable energy data were not available for all architectures. In particular, for RB and MBD, we rely on estimates derived from reported runtimes. Although this aligns with prior research indicating a relationship between runtime and energy-consumption~\cite{Bruce,Zhou18,Weber23}, these values remain approximations introduced to compare the various FDD architectures. For LLMs, we add further assumptions. Since LLMs are not used exclusively for building FDD, we allocate only a fraction of the overall LLM training cost to the diagnosis task. We also approximate the additional energy required for fine-tuning an existing LLM. As a result, the computational energy values should be interpreted as approximations.

Lastly, we do not account for all energy expenditures associated with developing and maintaining the different diagnosis methods. For example, feature selection, hyperparameter tuning, prompt design, rule creation, and creating models from first principles are not included in the reported training energy. While these activities may require substantial effort and computational resources, their energy demand is difficult to quantify consistently across the considered architectures.

\subsection{Generalizability}
First, the generalizability of our simulation results is limited by the HVAC subsystems considered. In particular, we focus on chiller and AHU applications. Consequently, the numerical results do not necessarily transfer directly to other HVAC systems or fault types. 

Furthermore, the relative benefit of a diagnosis approach depends on characteristics of the monitored systems. While we cover a selected range of parameters (e.g., building area), we do not consider other potentially relevant properties, including occupancy patterns, environmental conditions, or usage profiles. Accounting for such factors may change the relative benefit and ranking of the FDD methods considered. 

Finally, our current experiments rely on literature values obtained from specific implementations of the considered architectures and methods. A highly optimized implementation of a particular method may exhibit substantially different diagnostic performance or computational energy demand than the ranges we currently report. Therefore, the rankings observed in our simulations should not be interpreted as universal rankings of the FDD methods, but rather as an illustration of the underlying relationship: the preferred architecture depends on diagnostic performance, computational energy demand, and building scale.

\section{Conclusions}\label{sec:conclusion}

In this paper, we investigate a selection of software architectures for FDD in building systems. We introduce an energy-consumption model that relates the performance of an FDD architecture/method to the energy consumption of the monitored HVAC system. In an extended model, we also account for the energy-demand of model creation/learning and inference. We utilize the models in a set of initial simulation experiments to compare the various architectures and methods based on performance and energy expenditure values collected from prior literature.

The results indicate that for office buildings, the energy required for training and inference is negligible, since the energy demand of the HVAC system itself accounts for a substantial portion of the energy-consumption. In this case, fine-tuned LLMs provide the largest net energy savings. For single-family houses, the architecture's energy-demand becomes more relevant as the ecological footprint of heating and cooling is noticeably smaller. Nevertheless, fine-tuned LLMs also achieve the largest energy savings in this setup. Our experiments further show that inference frequency can influence the ranking of software architectures. 

Overall, there is no universally most energy-efficient FDD architecture. Rather, the superiority of an approach depends on the relationship between diagnostic performance, energy-consumption for training and inference, and the characteristics and operation of the monitored building. 

While our experiments allow us to draw some initial conclusions, the values collected from prior literature should be interpreted as estimates rather than universally precise figures. Future work should therefore evaluate the considered diagnosis architectures and approaches based on a common benchmark to collect diagnostic-performance  and energy values under the same conditions. Furthermore, we want to apply our energy-consumption models to real-world building data to assess the practical usefulness of our approach.


\section*{Acknowledgment}

The work presented in this paper has been supported by the FFG project Artificial Intelligence for Smart Diagnosis in Building Automation (ALFA) under grant 914932. 

\section*{Declaration of generative AI and AI-assisted technologies in the manuscript preparation process}
During the preparation of this work, the authors used Grammarly Pro (Model and exact version not publicly disclosed by the provider; accessed August 2026), Claude (Sonnet 5), and ChatGPT (GPT-5.6 Sol, OpenAI, San Francisco, CA, USA) in order to improve language, grammar, readability and the structure and organization of parts of the manuscript. After using these tools, the authors reviewed and edited the content as needed and take full responsibility for the content of the published article.

\newpage
\appendix
\section{Reviewed Literature on FDD in Chiller/AHU systems}
\label{app1}

\begin{landscape}
\scriptsize
\setlength{\tabcolsep}{1.2pt}
\renewcommand{\arraystretch}{1.12}

\setlength{\LTleft}{\fill}
\setlength{\LTright}{\fill}

\begin{longtable}{@{}
>{\RaggedRight\arraybackslash}p{0.9cm}
>{\RaggedRight\arraybackslash}p{1.3cm}
>{\RaggedRight\arraybackslash}p{4.3cm}
>{\RaggedRight\arraybackslash}p{3.0cm}
>{\RaggedRight\arraybackslash}p{1.8cm}
>{\RaggedRight\arraybackslash}p{2.0cm}
>{\RaggedRight\arraybackslash}p{1.4cm}
>{\RaggedRight\arraybackslash}p{2.0cm}
>{\RaggedRight\arraybackslash}p{0.7cm}
>{\RaggedRight\arraybackslash}p{0.7cm}
>{\RaggedRight\arraybackslash}p{0.7cm}
>{\RaggedRight\arraybackslash}p{0.7cm}
@{}}
\caption{Summary of target HVAC fault-detection/diagnosis methods reviewed by Chen et al.~\cite{chen2023review}. (NR = NOT REPORTED)}\label{tab:hvac_fdd_summary}\\
\toprule
\textbf{Paper ID} & \shortstack{\textbf{Chiller/}\\\textbf{AHU}} & \textbf{Summary} & \textbf{Method} & \textbf{TPR/Recall} & \textbf{TPR note} & \textbf{FPR} & \textbf{FPR note} & \textbf{TN} & \textbf{TP} & \textbf{FN} & \textbf{FP} \\
\midrule
\endfirsthead
\toprule
\textbf{Paper ID} & \shortstack{\textbf{Chiller/}\\\textbf{AHU}} & \textbf{Paper summary} & \textbf{Method} & \textbf{TPR/Recall} & \textbf{TPR note} & \textbf{FPR} & \textbf{FPR note} & \textbf{TN} & \textbf{TP} & \textbf{FN} & \textbf{FP} \\
\midrule
\endhead
\midrule
\multicolumn{12}{r}{\footnotesize Continued on next page}\\
\endfoot
\bottomrule
\endlastfoot
\cite{Namburu2007DataDriven} & Chiller & Centrifugal chiller (ASHRAE benchmark); nominal operation and 8 single-fault conditions at multiple severities and operating conditions. Generalized likelihood ratio test (GLRT) performs detection; SVM, principal component analysis (PCA), and partial least squares
(PLS) are used for fault classification. Reported: Classification Rate/Accuracy, average \% errors & SVM fault classifier (variant/kernel not reported) & NR & NR& NR & NR & NR & NR & NR & NR \\\hline

\cite{YAN14ARX}   & Chiller   &  ASHRAE RP-1043 90-ton chiller; five faults at 10--40\% severity; Compares direct SVM, regression-residual+SVM, auto-regressive moving-average model with exogenous variables (ARX)+MLP, and ARX+SVM    & SVM (RBF) & NR & NR & 0.20 & Paper reports false-alarm value 20.22 & NR & NR & NR & NR \\
 &  &  & regression-residual+SVM & NR & NR & 0.07 & Paper reports false-alarm value 7.38 & NR & NR & NR & NR \\
 &  &  & ARX+MLP & NR & NR & 0.02 & Paper reports false-alarm value 1.83 & NR & NR & NR & NR \\
 &  &  & ARX+ SVM & NR & NR & < 0.00 & Paper reports false-alarm value 0.11& NR & NR & NR & NR \\\hline

\cite{MULUMBA2015Robust} & AHU  & ASHRAE RP-1312 AHU; 11 faults. ReliefF feature selection, ARX parameters, SVM( RBF kernel +C-support vector classification (C-SVC)); Compares to raw-data ML baselines.  & SVM (RBF+C-SVC) + ReliefF + ARX  & 0.921 & Over all 11 fault classes & NR & NR & NR & NR & NR & NR \\
 &  &  & RBF network (NN)& 0.668 & Over all 11 fault classes  & NR & NR& NR & NR & NR & NR \\
&  &  & MLP & 0.481 & Over all 11 fault classes  & NR & NR & NR & NR & NR & NR \\
 &  &  & SVM (LibSVM library) & 0.785& Over all 11 fault classes  & NR & NR & NR & NR & NR & NR \\
 &  &  & DT& 0.794& Over all 11 fault classes  & NR & NR & NR & NR & NR & NR \\
&  &  & RF  & 0.817 & Over all 11 fault classes  & NR & NR & NR & NR & NR & NR \\\hline

\cite{jones2015fault} & AHU  & Simulation model (TRNSYS model) for experiments with four faults.  & RBS (Qualitative thresholds) & 0.11 &  & 0.00 &  & 7,242 & 216 & 1,675 & 0 \\
 &  &  & RBS (Quantitative) & 0.256 &  & 0.05 & & 6,936 & 484 & 1,407 & 389 \\
 &  &  & SVM (one-class SVM, radial-basis nonlinear kernel) & 0.31 &  & 0.28 &  & 7,304 & 587 & 1,304 & 21 \\
 &  &  & MLP  & 0.58 &  & 0.13 & & 6,409 & 1,105 & 780 & 916 \\
 &  &  & Single Fuzzy ART (NN) & 0.60& & 0.02 &  & 7,164 & 1,115 & 773 & 160 \\
 &  &  & LAPART (NN)& 0.47 &  & 0.04 &  & 7,044 & 889 & 1,002 & 279 \\\hline

\cite{YAN2020Generative} & Chiller   & ASHRAE RP-1043 90-ton chiller with seven faults at four severities;  Rebalancing with conditional Wasserstein generative adversarial networks (CWGAN); Compared across SVM, RF, DT, and LR. Reported Accuracy.    & SVM  & NR & NR & NR &NR& NR & NR & NR & NR \\
  &  &  & RF  & NR & NR & NR & NR & NR & NR & NR & NR \\
  &  &  & DT  & NR & NR & NR & NR & NR & NR & NR & NR \\
 &  &  & LR & NR & NR & NR & NR & NR & NR & NR & NR \\\hline

\cite{TAHERI2021FaultDetection}   & AHU     & Single- and multi-zone VAV AHUs from LBNL FLEXLAB; Six faults plus normal operation. Seven deep recurrent neural networks (DRNN) configurations are optimized and compared with RF and gradient boosting; Heat Map representation of confusion matrix, no values. Reported, Precision; Recall; (multiclass aggregate, averaging not specified); Accuracy; F1-Score; ROC  & Stacked and deep transition output combined (S,DTO)-DRNN (SZ-AHU) & 0.92 & multiclass& NR & NR & NR & NR & NR & NR \\
&  &  & S,DTO-DRNN (MZ-AHU) & 0.88 & multiclass& NR & NR & NR & NR & NR & NR \\
 &  &  & RF  (SZ-AHU)& 0.89 & multiclass & NR & NR & NR & NR & NR & NR \\
 &  &  & RF(Gradient boosting) (SZ-AHU) & 0.81 & multiclass & NR & NR& NR & NR & NR & NR \\
  &  &  & RF  (MZ-AHU)& 0.80 & multiclass & NR & NR & NR & NR & NR & NR \\
 &  &  & RF (Gradient boosting) (MZ-AHU)& 0.75 & multiclass & NR & NR& NR & NR & NR & NR \\

  &  &  & & &&&&&&&\\ \hline
\cite{YAN2020Unsupervised}   & AHU  & ASHRAE RP-1312 data from two AHUs; Six faults. Training uses 5/10/20/30/40 faulty samples per fault type plus 7200 normal samples; Conditional Wasserstein Generative Adversarial Network with Ensemble Learning Quality Control Protocol (CWGAN-ELQCP) rebalancing is compared with conventional RF, SVM, MLP, and DT  & RF (conventional) & 0.167, 0.182, 0.176, 0.169, 0.169 & Values correspond to 5, 10, 20, 30, 40 faulty training samples per fault type & NR & NR & NR & NR & NR & NR \\
&  &  & RF (proposed) &  0.979, 0.974, 0.970, 0.982, 0.999 & Values correspond to 5, 10, 20, 30, 40 faulty training samples per fault type & NR & NR & NR & NR & NR & NR \\
&  &  & SVM (conventional) & 0.186, 0.163, 0.179, 0.172, 0.177 & Values correspond to 5, 10, 20, 30, 40 faulty training samples per fault type & NR & NR & NR & NR & NR & NR \\
&  &  & SVM (proposed) &  0.995, 0.995, 0.997, 0.996, 0.999  & Values correspond to 5, 10, 20, 30, 40 faulty training samples per fault type& NR & NR & NR & NR & NR & NR \\
 &  &  & MLP (conventional)  & 0.168, 0.175, 0.182, 0.181, 0.183 & Values correspond to 5, 10, 20, 30, 40 faulty training samples per fault type & NR & NR & NR & NR & NR & NR \\
  &  &  & MLP (proposed)  & 0.975, 0.962, 0.971, 0.984, 0.998 & Values correspond to 5, 10, 20, 30, 40 faulty training samples per fault type & NR & NR & NR & NR & NR & NR \\
 &  &  & DT  (conventional)  & 0.168, 0.168, 0.167, 0.166, 0.173& Values correspond to 5, 10, 20, 30, 40 faulty training samples per fault type  & NR & NR & NR & NR & NR & NR \\
  &  &  & DT (proposed)   & 0.954, 0.917, 0.944, 0.912, 0.999& Values correspond to 5, 10, 20, 30, 40 faulty training samples per fault type  & NR & NR & NR & NR & NR & NR \\\hline

\cite{HAN2020Ensemble}   & Chiller  & ASHRAE RP-1043 90-ton centrifugal chiller with seven gradual faults.  12,000 samples (8,000 train/validation; 4,000 test) with grid search and 10-fold CV; Compares optimized SVM/RF and an ensemble; exact confusion counts are available.   & SVM (RBF kernel)  & >0.99 & Computed & 0.01 & Computed & 484 & 3498 & 13 & 5 \\
 &  &  & RF & >0.99& Computed & < 0.01 & Computed & 487 & 3509 & 2 & 2 \\
   &  &  & & &&&&&&&\\
     &  &  & & &&&&&&&\\
          &  &  & & &&&&&&&\\
       &  &  & & &&&&&&&\\ \hline

\cite{LI2016FaultDetection}   & Chiller   & ASHRAE RP-1043 90-ton water-cooled chiller; 7 faults at 4 severities plus normal data; 8 training sizes and raw versus preprocessed; Compares Tree-structured Fault Dependence Kernel (TFDK), SVM (RBF kernel), DT and LR&  SVM (TFDK)  -- small sample size & 0.9983 & Computed &  0.1025 & Computed & 359 & 11181 & 19& 41\\
 &  &  & SVM (TFDK)  -- large sample size & 1.0000 & Computed &  0.0000 & Computed & 400 & 11200 & 0& 0\\
 &  &  & SVM (RBF kernel) -- small sample size & 0.9908 & Computed & 0.1225& Computed & 351 & 11097 & 103& 49 \\
  &  &  & SVM (RBF kernel) -- large sample size &  0.9995 & Computed & 0.0250 & Computed & 390 & 11194 &  6 & 10 \\
 &  &  & DT  & NR & NR & NR & NR & NR & NR & NR & NR \\
 &  &  & LR  & NR & NR & NR & NR & NR & NR & NR & NR \\\hline

\cite{LI2021ANovel}  & Chiller  & ASHRAE RP-1043 90-ton chiller with 7 faults at 4 severities. Compares Semi-GAN with MLP and SVM baselines; Reports only  accuracy  & Semi-GAN (NN)& NR & NR & NR & NR & NR & NR & NR & NR \\
&  &  & MLP (1 layer) & NR & NR & NR & NR & NR & NR & NR & NR \\
 &  &  & MLP (2 layers) & NR & NR & NR & NR & NR & NR & NR & NR \\
&  &  & Least-squares SVM (LS-SVM) & NR & NR & NR & NR & NR & NR & NR & NR \\
&  &  & Semi-SVM-based  method & NR & NR & NR & NR & NR & NR & NR & NR \\\hline

\cite{WANG2021FaultDiagnosis} & Chiller & ASHRAE RP-1043 90-ton chiller with 7 faults at 4 severities. Hybrid method using  model-based methods and Bayesian network. Reports accuracy only. & -- &  &  &  &  &  &  &  &  \\\hline

\cite{khr27}   & Chiller   & Empirical comparison of rule-based, classification, and outlier-detection approaches for HVAC fault detection using a simulated chiller benchmark. 7 faults; 5 fold CV  & RBS & 0.3851 & Mean recall over 5 CV folds; fault = positive class. & NR & NR & NR & NR & NR & NR \\
&  & & DT & 0.7937 & Mean recall over 5 CV folds; fault = positive class. & NR & NR & NR & NR & NR & NR \\
&  & & DT(feature-selected)& 0.7757 & Mean recall over 5 CV folds; fault = positive class. & NR & NR & NR & NR & NR & NR \\
&  & & RF & 0.7744 & Mean recall over 5 CV folds; fault = positive class. & NR & NR & NR & NR & NR & NR \\
&  & & RF (feature-selected) & 0.7782 & Mean recall over 5 CV folds; fault = positive class. & NR & NR & NR & NR & NR & NR \\
&  & & LR  (100 iterations) & 0.4389 & Mean recall over 5 CV folds; fault = positive class. & NR & NR & NR & NR & NR & NR \\
&  & & LR (1000 iterations) & 0.3780 & Mean recall over 5 CV folds; fault = positive class. & NR & NR & NR & NR & NR & NR \\
&  & & MLP & 0.3747 & Mean recall over 5 CV folds; fault = positive class. & NR & NR & NR & NR & NR & NR \\
&  & & MLP (feature-selected) & 0.4852 & Mean recall over 5 CV folds; fault = positive class. & NR & NR & NR & NR & NR & NR \\

\end{longtable}
\end{landscape}

\begin{landscape}
\scriptsize
\setlength{\tabcolsep}{1.2pt}
\renewcommand{\arraystretch}{1.12}

\setlength{\LTleft}{\fill}
\setlength{\LTright}{\fill}

\begin{longtable}{@{}
>{\RaggedRight\arraybackslash}p{0.9cm}
>{\RaggedRight\arraybackslash}p{1.3cm}
>{\RaggedRight\arraybackslash}p{4.3cm}
>{\RaggedRight\arraybackslash}p{3.0cm}
>{\RaggedRight\arraybackslash}p{1.8cm}
>{\RaggedRight\arraybackslash}p{2.0cm}
>{\RaggedRight\arraybackslash}p{1.4cm}
>{\RaggedRight\arraybackslash}p{2.0cm}
>{\RaggedRight\arraybackslash}p{0.7cm}
>{\RaggedRight\arraybackslash}p{0.7cm}
>{\RaggedRight\arraybackslash}p{0.7cm}
>{\RaggedRight\arraybackslash}p{0.7cm}
@{}}
\caption{Summary of additional sources considered for TPR/FPR aggregation. (NR = NOT REPORTED)}\label{tab:hvac_fdd_summary_additional_sources}\\
\toprule
\textbf{Paper ID} & \shortstack{\textbf{Chiller/}\\\textbf{AHU}} & \textbf{Summary} & \textbf{Method} & \textbf{TPR/Recall} & \textbf{TPR note} & \textbf{FPR} & \textbf{FPR note} & \textbf{TN} & \textbf{TP} & \textbf{FN} & \textbf{FP} \\
\midrule
\endfirsthead
\toprule
\textbf{Paper ID} & \shortstack{\textbf{Chiller/}\\\textbf{AHU}} & \textbf{Paper summary} & \textbf{Method} & \textbf{TPR/Recall} & \textbf{TPR note} & \textbf{FPR} & \textbf{FPR note} & \textbf{TN} & \textbf{TP} & \textbf{FN} & \textbf{FP} \\
\midrule
\endhead
\midrule
\multicolumn{12}{r}{\footnotesize Continued on next page}\\
\endfoot
\bottomrule
\endlastfoot
\cite{CUI2005986} & Chiller & ASHRAE RP-1043 centrifugal chiller. Model-based with adaptive fault detection threshold estimator. Five fault types over four severity levels. & MBD & 0.51, 0.62, 0.20, 0.69, 1.00 & Detection rates for: reduced evaporator-water flow, refrigerant leakage, excess oil, condenser fouling, and non-condensables & NR & The paper reports a ``false alarm rate'', but its denominator is not sufficiently clear to treat it as FPR. & NR & NR & NR & NR \\\hline

\cite{Padmanabh21} & Chiller & Real-building BMS chiller data; Logistic-circuit decision tree to determine patterns predicts alarms. & Decision tree & 0.76 & Reported & 0.000385 & Computed & 83061 & 162 & 51 & 32 \\\hline

\cite{mchugh2019fault} & AHU & UT Austin main-campus BAS data; chilled-water leakage classification; final tuned model. & Decision tree (chilled water) & 0.990361 & Computed & 0.012251 & Computed & 13061 & 8322 & 81 & 162 \\
& & & Decision tree (steam) & 0.966288 & Computed & 0.026026
& Computed & 50745 & 40272 & 1405 & 1356\\\hline

\cite{langerThesis} & Chiller & EnergyPlus-Modelica SDAHU dataset; single-fault binary classification & LLM (fine-tuned) - Fine-tuned DistilBERT & 0.98 & Derived from reported FNR & 0.02 & Reported& NR & NR & NR & NR \\
 & AHU & EnergyPlus-Modelica SDAHU dataset; single-fault binary classification & LLM (fine-tuned) - Fine-tuned DistilBERT & 0.99 & Derived from reported FNR & 0.01 & Reported& NR & NR & NR & NR \\
& AHU & & LR & NR & NR & 0.00 & Reported FPR & NR & NR & NR & NR\\\hline

\cite{Islam24} & AHU & LBNL SZVAV (experimental) and SZCAV (simulated); binary fault classification in a hospital-building FDD study.& LR & 0.9164 & Computed & 0.5303 & Computed & 674 & 5439 & 496 & 761 \\\hline

\cite{ZHANG2025124378} & AHU & ASHRAE RP-1312 containing 12 AHU fault classes. GPT-3.5 is fine-tuned using a labeled-data-supervised fine-tuning method. & LLM (Original GPT-3.5) & 0.25 & Aggregated recall reconstructed & 0.13 & FPR reconstructed from fault-free class accuracy & NR & NR & NR & NR \\
 &  &  & LLM (tuned) - Fine-tuned GPT-3.5 (Version 4) & 1.00 & All fault classes have 100\% diagnosis accuracy. & 0.00 & All fault-free classes have 100\% accuracy. & NR & NR & NR & NR\\

\end{longtable}
\end{landscape}

\bibliographystyle{plain}
\bibliography{aloha,mybibfile,references} 

\end{document}